\documentclass[12pt, preprint]{aastex}
\usepackage{epsf}
\usepackage{times}
\newcommand{\etal}{{et al}\/.}
\newcommand{\hh}{^{\rm h}}
\newcommand{\mm}{^{\rm m}}
\begin{document}
\slugcomment{Draft of \today}
\shorttitle{Centaurus A jet}
\shortauthors{M.J.\ Hardcastle \etal}
\title{Radio and X-ray observations of the jet in Centaurus A}
\author{M.J.\ Hardcastle and D.M.\ Worrall}
\affil{Department of Physics, University of Bristol, Tyndall Avenue,
Bristol BS8 1TL, UK}
\and
\author{R.P.\ Kraft, W.R.\ Forman, C.\ Jones and S.S.\ Murray}
\affil{Harvard-Smithsonian Center for Astrophysics, 60 Garden Street, Cambridge, MA~02138, USA}
\begin{abstract}
We present new, high dynamic range VLA images of the inner jet of the
closest radio galaxy, Centaurus A. Over a ten-year baseline we detect
apparent sub-luminal motions ($v \sim 0.5c$) in the jet on scales of
hundreds of pc. The inferred speeds are larger than those previously
determined using VLBI on smaller scales, and provide new constraints
on the angle made by the jet to the line of sight if we assume
jet-counterjet symmetry. The new images also allow us to detect faint
radio counterparts to a number of previously unidentified X-ray knots
in the inner part of the jet and counterjet, showing conclusively that
these X-ray features are genuinely associated with the outflow.
However, we find that the knots with the highest X-ray to radio flux
density ratios do not have detectable proper motions, suggesting that
they may be related to standing shocks in the jet; we consider some
possible internal obstacles that the jet may encounter. Using new,
high-resolution {\it Chandra} data, we discuss the radio to X-ray
spectra of the jet and the discrete features that it contains, and
argue that the compact radio and X-ray knots are privileged sites for
the {\it in situ} particle acceleration that must be taking place
throughout the jet. We show that the offsets observed between the
peaks of the radio and X-ray emission at several places in the Cen A
jet are not compatible with the simplest possible models involving
particle acceleration and downstream advection together with
synchrotron and expansion losses.
\end{abstract}

\maketitle
\section{Introduction}

In the years since the launch of {\it Chandra} it has become apparent
that many FRI radio galaxies and BL Lac objects have kpc-scale X-ray
jets \citep*[e.g.,][and refs therein]{wbh01,hk02}. On the basis of the
continuity of the radio/optical/X-ray spectrum, it has been argued in
several cases \citep*[e.g.,][]{hbw01} that the
emission mechanism is synchrotron radiation. In this case, the X-rays
are giving us information on a population of extremely energetic
electrons, with random Lorentz factors ($\gamma$) as high as $10^8$.
It may even be the case that {\it all} kpc-scale FRI jets (including
those of the BL Lac objects, the unification partners of FRI radio
galaxies) are synchrotron X-ray sources at some level, with the
high-energy particle acceleration being linked to the strong
deceleration that the jet is known to undergo on these scales.

Although the overall morphological agreement between the radio,
optical (where observed) and X-ray jets is generally good, reinforcing
our confidence in a synchrotron model, there are often significant
differences in detail: particularly notable are the tendency for the
inner parts of the jet to have a higher X-ray to radio flux density
ratio, and the offsets between the peak positions of X-ray and radio
knots, which are typically in the sense that the X-ray knot's position
is closer to the core \citep{hbw01,hwbl02}. To date there
has been no particularly satisfactory explanation for these
observations. Of course, we should not expect to see a detailed
agreement between the radio and X-ray images, if the X-ray emission is
synchrotron: the loss timescale for the radio-emitting electrons is of
the order of hundreds of thousands of years, assuming a magnetic field
strength close to the equipartition value, while the loss timescale
for an X-ray-emitting electron in the same field is of the order of
tens of years. In other words, the X-ray-emitting electrons are
telling us where particles are being accelerated {\it now}, while the
radio-emitting electrons tell us about the time-averaged particle
acceleration, combined with the effects of downstream motion (i.e.
motion away from the core). Arguments of this kind have been used to
explain the observed offsets, but in most sources it is hard to make
them quantitative because of the limited spatial resolution of {\it
Chandra}.

Centaurus A, at a distance of 3.4 Mpc \citep{i98}, is the closest
radio galaxy: $1''$ corresponds to 17 pc. Cen A is therefore the only
source where the spatial size corresponding to {\it Chandra}'s
sub-arcsecond angular resolution is comparable to the energy-loss
travel distance of the X-ray emitting electrons (assuming moderately
relativistic bulk motion). An understanding of the processes going on
in Cen A is critical if we are to understand the X-ray jets in other,
more distant FRI sources. In earlier papers \citep[][\ {~[hereafter
K02]}]{kfjk00, kfjm02} we have presented {\it Chandra} and radio
observations of Cen A, showing that the radio/X-ray relationship is
complex. In this paper we present new, high-dynamic-range radio data
and new, high-spatial-resolution X-ray imaging spectroscopy, which
together shed new light on the dynamics and acceleration processes in
the jet and counterjet.

J2000.0 co-ordinates are used throughout this paper, and the spectral
index $\alpha$ is defined in the sense that $S_\nu \propto \nu^{-\alpha}$.

\section{Radio observations and data reduction}
We observed Cen A using the NRAO Very Large Array (VLA) in A and B
configurations at 8.4 GHz in 2002. The source had previously been
observed (PI Burns) in the A, B, C and DnC VLA configurations in
1990/91; we presented earlier images from these observations in K02.
In this paper we make use only of the A and B-configuration data from
the earlier observations. Observational details for both epochs are
given in Table \ref{vlaobs}. Apart from the use in 2002 of a closer
phase calibrator, and the narrow bandwidth used in the 1991 B-array
observations, the 1991 and 2002 observations were very similar.
3C\,286 was used as the flux calibrator in all observations, and in
all cases Cen A was observed for essentially the whole time permitted
by the elevation limits of the VLA antennae.

\begin{deluxetable}{lllllll}
\tablecaption{VLA observations}
\tablehead{Date&Program ID&Configuration&Time on&Frequencies&Bandwidth&Phase\\&&&source (h)&(GHz)&(MHz)&calibrator\\}
\startdata
1991 Jul 02&AB587&A&2.1&8.415, 8.455&50&1337$-$129\\
1991 Nov 09&AB587&B&2.1&8.434, 8.484&12.5&1337$-$129\\
2002 Mar 03&AH764&A&2.9&8.435, 8.485&50&1316$-$336\\
2002 Jul 12&AH764&B&2.9&8.435, 8.485&50&1316$-$336\\
\enddata
\label{vlaobs}
\end{deluxetable}

The data reduction was initially carried out in a standard manner
using {\sc aips}. As reported in K02, the individual observations from
1991 were flux- and phase-calibrated, flagged, and then
self-calibrated in phase and amplitude, starting with a point-source
model in the case of the A-array data, to give images with respectable
but limited dynamic range, $\sim 10^4:1$. A more
realistic measure of the image quality, on-source peak to off-source
peak, was 2000:1; artefacts around the strong core gave rise to
structure in the noise.

The 2002 data were reduced in a very similar way, and immediately gave
significantly better results\footnote{The most plausible explanation
for the superior quality of the 2002 data is that it is due to changes
in the VLA's calculation of correlation coefficients, which were
implemented in 1998: \citet*{tup02}, section 3.10.}. But these images
were still limited by artefacts around the core, which showed up at
around 10 times the r.m.s. noise level. Since it was possible to make
an image that contained all the flux at A-array, we then attempted
{\it baseline-based} self-calibration, using the {\sc aips} task {\it
blcal} to generate a set of baseline corrections from the image that
could then be applied directly to the data. Because of the danger of
forcing the data to match the model using this method, we determined
only a single set of baseline-based corrections for all times present
in the dataset. The corrections were then applied to the A-array
dataset using the {\sc aips} task {\it split}. The result, after deep
cleaning using the {\it imagr} task, was an image with an off-source
noise level of 50 $\mu$Jy beam$^{-1}$, and no obvious core-related
artefacts. The noise level was still some way above the expected
thermal value ($\sim 10$ $\mu$Jy beam$^{-1}$), but corresponded to a
dynamic range of 120,000:1, among the highest ever achieved with the
VLA in continuum mode. By examining the difference between the
baseline-corrected and non-baseline-corrected maps, we verified that
no significant changes in source structure had been forced by the
baseline-based calibration.

It was not possible to apply the same method to the 1991 A-array
dataset, because an adequate model (representing all the flux visible
in the $uv$ dataset) could not be made from the image derived from
self-calibration. Instead, we amplitude- and phase-calibrated the 1991
dataset, and then determined baseline corrections, using an image made
from the 2002 data (after using {\it uvsub} to correct for the effects
of core variability). The result was an image with an off-source r.m.s.
noise of 98 $\mu$Jy beam$^{-1}$, not as good as that in the 2002 map,
but still acceptable. A comparison of the maps before and after this
process showed that no significant changes to the source structure had
been forced by the cross-calibration of the datasets, in the sense
that subtraction of the maps showed no structure that could not be
attributed to noise in the non-baseline-corrected dataset.

Finally, the B-array data were cross-calibrated and baseline
self-calibrated in a very similar way (although in this case we found
that some core-related artefacts in the 1991 data could only be
removed by a time-varying baseline-based calibration). It was then possible
to combine the individual A and B-array datasets to produce images
representing both compact and relatively extended structure, or to
combine all four datasets to produce an image with better sampling and
(presumably) fidelity than could be provided by the individual epochs
of observation. We use this multi-epoch dataset when comparing with
the X-ray data discussed later in the paper.

Imaging during the calibration process was carried out using the {\it
imagr} task only. After calibration was complete we experimented with
using the maximum-entropy deconvolution routine {\it vtess}. Using
standard `hybrid' mapping techniques to remove the flux from the
bright core before applying {\it vtess}, and convolving the
maximum-entropy images with the same Gaussian beam as that fitted by
{\it imagr} to the $uv$ data, we obtained images which appeared
smoother than the {\it clean}-based ones, although they suffered from
a slight positive bias. Subtraction of the {\it vtess} and {\it imagr}
images revealed striping in the extended parts of the residual image,
which we attribute to the known instabilities in the {\it clean}
algorithm. As this striping has an amplitude of up to 50\% of the
total in the faint extended emission, it could potentially have
serious effects on the measurements of positions of faint or extended
features in the jet. Accordingly, we use {\it vtess}-derived maps for
positional measurements. However, we note that bright knots appear
identical in the {\it imagr} and hybrid maps (i.e., image subtraction
gives residuals close to zero), and their positions as determined
using {\it jmfit} change by only a few mas if {\it vtess} rather than
{\it imagr} is used, so that we do not believe that the deconvolution
method has a significant effect on the positions estimated for bright
jet features.

There are artefacts around the core in all polarization (Stokes $Q$
and $U$) maps (which were all made using {\it imagr}). We attribute
the artefacts to the limited accuracy of the correction for the
`leakage' terms (determining the amount of unpolarized flux which
appears in the polarization channels) carried out by the task {\it
pcal}. The core appears polarized at about the 0.2\% level, consistent
with the expected performance of {\it pcal} (particularly as the
parallactic angle coverage of the phase calibrator was limited by the
short observing time) but enough to cause the observed artefacts.
Since the artefacts are different from dataset to dataset,
polarization images based on more than one dataset are particularly
badly affected, and so we do not present maps made from such images
here.

The initial calibration of the A-array data with a point source model
caused us to lose the absolute astrometry of the images (which, as
discussed in K02, was not very accurate in the case of the 1991 data
in any case). To correct for this, we followed K02 and set the
positions of the cores in all observations to accurate values derived
from an archival 8-GHz Australia Telescope Compact Array (ATCA)
observation which had not been calibrated in this way. Our adopted
radio core position is $13\hh 25\mm 27\fs 609$, $-43\degr 01'
08\farcs91$. Because our two images are referenced at the core, we are
implicitly assuming that in the search for proper motion described
below (\S\ref{motion}) the core is stationary. However, as we shall
see, the full range of detected knot motions cannot possibly be
attributed to changes in core structure.

Radio beam sizes quoted in what follows are the major and minor full
widths at half maximum (FWHM) of the restoring or convolving
elliptical Gaussians, derived in the usual way by fitting to the
center of the dirty beam. Because of the extreme southern declination
of Cen A, the long axis of the restoring beam is always oriented
within a few degrees of the north-south direction, and so its position
angle is not quoted.

\section{X-ray observations}

The X-ray observations were taken on 2002 September 03 with {\it
Chandra}, as part of the High Resolution Camera (HRC) guaranteed time
program, and were made with the jet on the back-illuminated (S3) chip
of the ACIS array for maximal sensitivity to soft photons. The active
nucleus was at the aim point (whereas it was 4--5' off-axis in the
earlier {\it Chandra} observations: K02), so that we have in these
observations, for the first time, a combination of sub-arcsecond
resolution and good spectral sensitivity for the whole of the inner
jet; the PSF in the inner jet is a factor $\sim 2$ smaller than in the
previous ACIS observations. The roll angle of the satellite was chosen
so that the frame transfer streak from the heavily piled-up nucleus of
Cen A is perpendicular to the jet direction; it also places the X-ray
arc known to lie around the southern inner lobe (K02; \citealt{kvfj03}) on the front-illuminated S2 chip, giving us a sensitive and
high-resolution image of that feature. In addition, there is some weak
evidence for emission from the edge of the northern inner lobe. We
will discuss these features in more detail elsewhere. Here we
concentrate on the jet, all of whose X-ray emission lies on the S3
chip.

We inspected the background of the observations as a function of time
and found no strong variations in count rate: the background appeared
to be at the expected level. Accordingly, no time filtering was
carried out. The effective exposure time was 45,182 s. To maximize the
effective resolution, we generated a new level 2 events file without
the 0.5-pixel randomization applied by the standard pipeline. We also
processed the data to remove the effects of `streaking' on the S4
chip. As reported by K02, we had manually adjusted the aspect solution
of the earlier {\it Chandra} observations to bring them in line with
known optical positions. No such adjustment was necessary for the new
observations. The alignment between the radio and X-ray core positions
is excellent.

We use the energy range 0.4--7.0 keV for all spectroscopy
in what follows, except where otherwise stated. For imaging we mostly
use the band 0.4--2.5 keV, as this removes much of the strong emission
from the heavily absorbed X-ray nucleus without compromising the soft
emission from the jet.

\section{Results}

\subsection{Radio emission}

\subsubsection{Knots and proper motions}
\label{motion}

\begin{figure*}
\epsscale{0.9}
\plotone{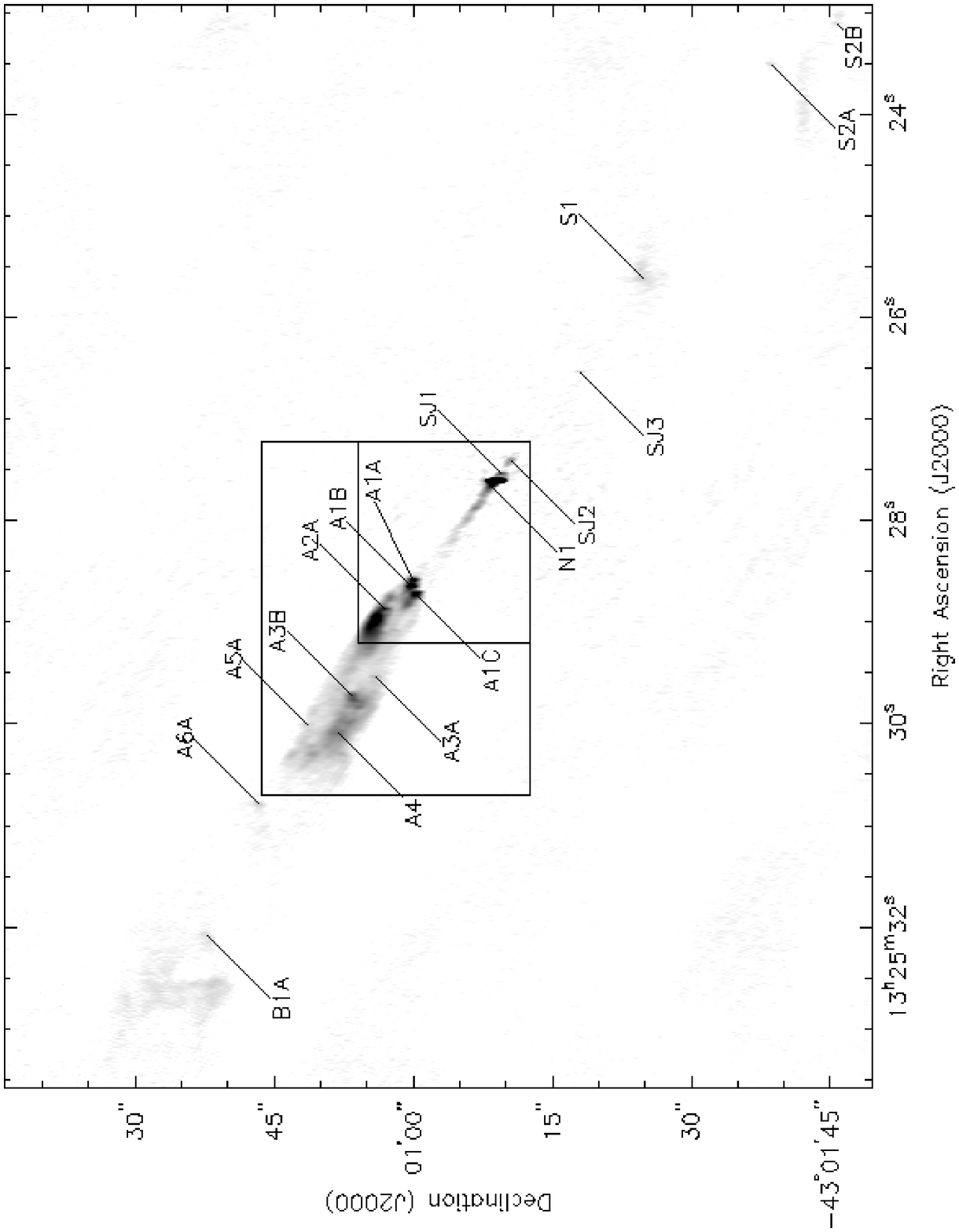}
\epsscale{1.0}
\caption{Radio knots in the jet and counterjet of Cen A. The map shown
  is the maximum-entropy 2002 A-array map, with $0\farcs 76 \times
  0\farcs 20$ resolution. Black is 10 mJy beam$^{-1}$. The transfer
  function is non-linear to allow low-surface-brightness structure to
  be seen. Knots are labeled according to a modified version of the
  notation of \cite{cbn92}. The initial letter(s) and number denote
  the large-scale feature of which the knots form a part: the final
  letter (where present) distinguishes between sub-knots. N1 is the
  bright radio knot in the inner jet close to the nucleus seen by
  \citeauthor{cbn92}. Boxes indicate regions of the jet that will be
  mapped in more detail in later Figures.}
\label{radio}
\end{figure*}

The high-dynamic-range images of the jet and counterjet (Fig.\
\ref{radio}) give us two important pieces of information. Firstly, a
number of compact faint radio knots are detected, including (for the
first time) several clearly compact features in the large-scale
counterjet region. (The inner counterjet knots denoted SJ1 and SJ2, and the
large-scale features S1 and S2, had already been detected in the maps
of \citeauthor*{cbn92} [\citeyear{cbn92}]). The new knots (A2A, A3A/B, A5A,
B1A, SJ3, S2A, S2B) are all faint (with flux densities of at most a
few mJy at 8.4 GHz) but are unambiguously detected in both VLA imaging
epochs. As we shall see below (\S \ref{RX}), several of the new faint
radio knots are associated with comparatively bright X-ray emission.

Secondly, mapping the difference between the two epochs (or even
simply blinking between the two maps) reveals that some, but not all,
of the features in the jet have detectable proper motions along the
jet. The most obvious motion is that of A1B, the middle knot in the
bright base-knot complex, which has an apparent motion on the sky over
the 11-year baseline of $0\farcs 101 \pm 0\farcs 001$ in a position
angle of 62\degr\ (defined in the sense north through east). The error
quoted here is based on the errors returned by the {\sc aips} Gaussian
fitting task {\it jmfit}, fitting to both source and background, and
so is likely to be optimistic, since it does not take into account
systematic uncertainties due to, for example, the choice of fitting
region: nevertheless, the proper motion is clearly detected. Further
down the jet, the large-scale regions A2 and A3/4 appear to be moving
{\it coherently} downstream. The apparent speed in these regions is
hard to quantify, because there are few well-defined bright knots, but
a measurement of the motion of the knot A3B, the brightest compact
feature in this region, gives a proper motion of $0\farcs06 \pm
0\farcs01$. By contrast, most of the {\it compact} features in the jet
and counterjet (A1A, A1C, A2A, A3A, A5A, B1A, SJ1, SJ2, SJ3, S2A, S2B)
had no detectable proper motion within the errors, although the
accuracy of positions determined by fitting Gaussians to these
features is low in the case of the fainter knots. In addition to the
proper motions, the inner knot (A1A) has varied significantly
(increasing in flux by 10\%) over the epoch of observations, while the
extended emission downstream of A1C appears to have become fainter.

\begin{figure*}
\plotone{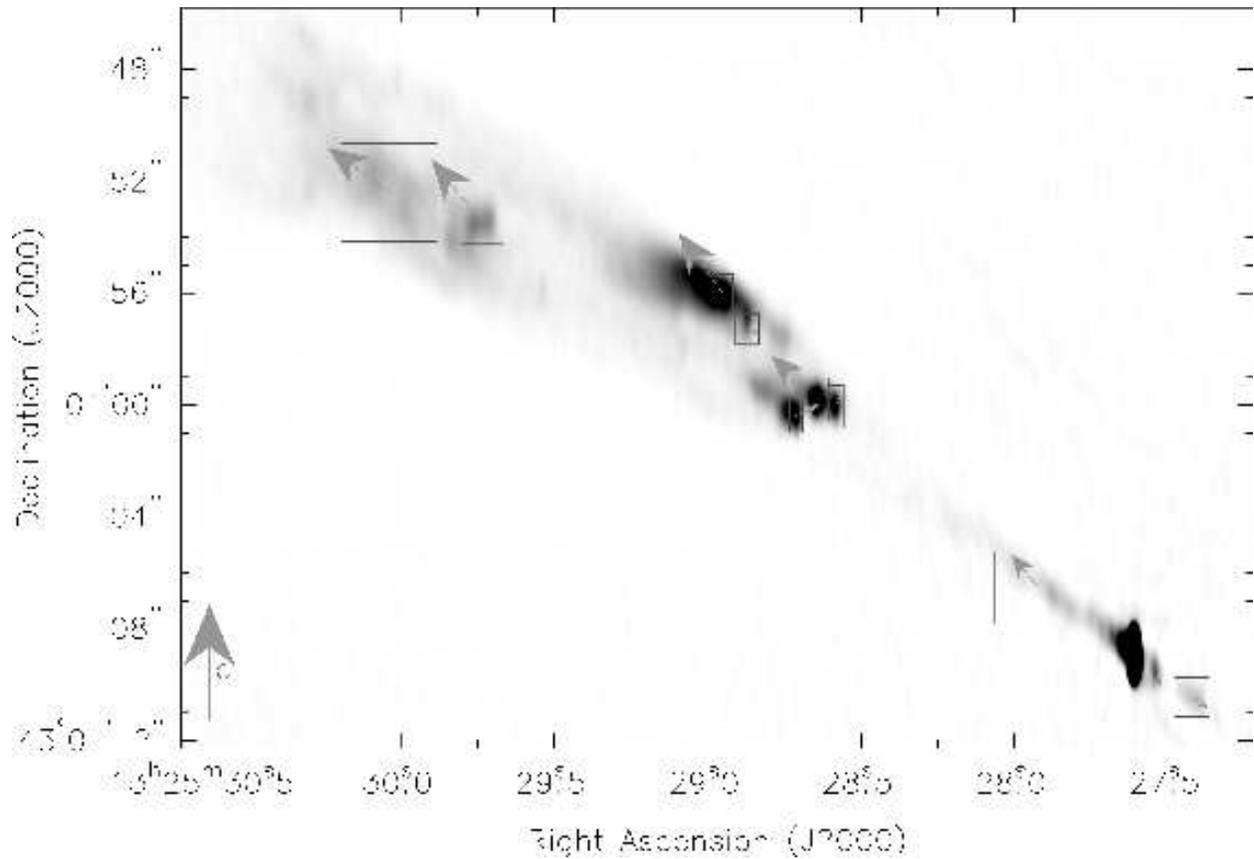}
\caption{Motions in the jet between 2002 and 1991. The greyscale shows
  the 2002 A-configuration image, convolved to a resolution of
  $0\farcs 77 \times 0\farcs 20$. Boxes indicate regions within which
  this image was compared with the 1991 image, convolved to the same
  resolution. Vectors show the best-fitting offsets between the two
  maps for each sub-image, exaggerated by a factor 20 for visibility.
  The vector in the bottom left-hand corner shows the offset that
  would be observed for a feature with an apparent speed of $c$. Note
  that only the four largest apparent motions are significantly
  detected; see the text for more details. This image shows the region
  of the jet denoted by the outer box in Fig.\ \ref{radio}.}
\label{walker}
\end{figure*}

To confirm the reality of the apparent motions we used a version of
the least-squares method discussed by \cite{w97}, which involves
shifting selected sub-regions of the jet so as to give the best match
between the two epochs. The regions used in this method and the
resulting best-fitting shifts are plotted in Fig.\ \ref{walker}. The
difficulties in applying this method to our data come from differences
in background structure (due to slightly different short-baseline $uv$
coverage) coupled with real changes in knot structure, such as those
seen in A1C. The best-fitting shift for the feature with the highest
SNR, A1B, is in good agreement (within the errors) between the two
methods: Walker's method gives it a motion of $0\farcs 12 \pm
0\farcs03$ (errors are $1\sigma$, derived from the least-squared fits
and based on estimates of the on-source noise). The motions determined
for the compact features A1A and A2A are consistent with zero. A1C's
apparent backward motion, which is formally marginally significant, is
a result of the changes in the knot structure discussed above. Further
out, Walker's method gives somewhat higher estimated speeds for the A3
region, $0\farcs 14\pm 0\farcs03$. Although the motions in the inner
jet and counterjet are not formally significant, they are plotted
because of the suggestive directions of the best-fitting shift vector;
it will be of interest to see whether these become significant in
future monitoring.

The motion of knot A1B corresponds to an apparent speed on the sky of
$0.51c$, based on the {\sc jmfit} results, while the apparent speed in
region A3 is between $0.3$ and $0.7c$ depending on the method used. We
are therefore observing subluminal proper motions in the kpc-scale jet
of Cen A, with apparent speeds higher than those observed in the
parsec-scale jet and counterjet \citep{tjrt98}. If the speeds are
taken to represent the bulk speeds of the source, this would imply jet
{\it acceleration} on scales between $\sim 1$ and 250 pc. It is more
plausible that the motions of the parsec-scale knots (and possibly
also of knot A1B) do not trace the bulk fluid flow in the jet, and in
fact \citeauthor{tjrt98} argue that the variability of sub-components
of the parsec-scale jet implies speeds $>0.45c$. A similar trend, in
the sense that apparent speed increases with distance from the
nucleus, has been observed in VLBI studies of some core-dominated
objects \citep{howr01}.

\begin{figure*}
\plotone{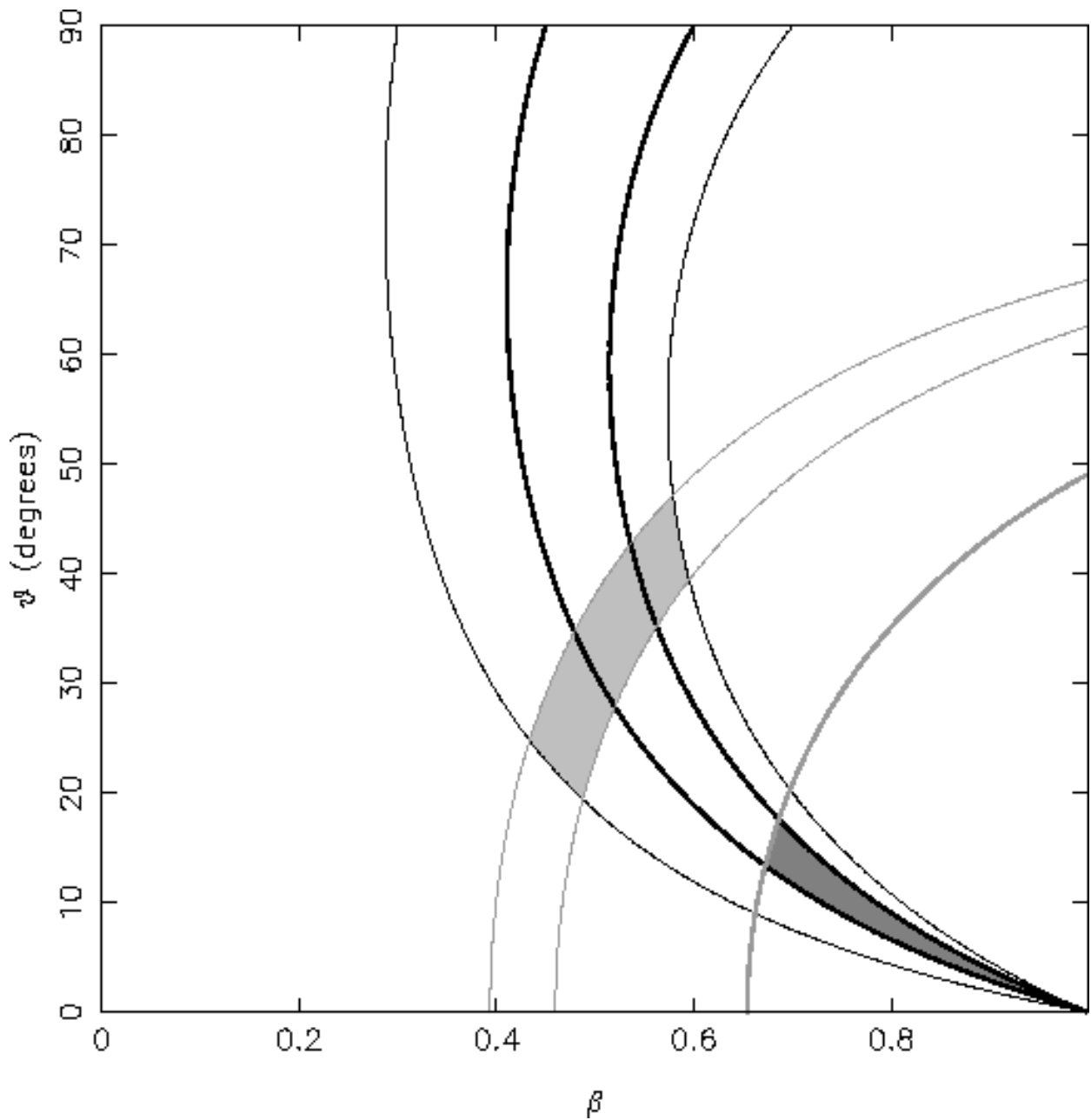}
\caption{Constraints on the permitted values of the bulk speed and
  angle to the line of sight from the jet-counterjet ratio and
  apparent motions in the A1B and A3 regions of the Cen A jet. Thick
  lines represent constraints on A1B, thin lines show constraints on
  A3. The gray lines show the constraints from sidedness, the black
  lines show the constraints due to apparent motions), and the
  intersection between the regions (shaded in light gray) shows the
  permitted regions of parameter space for the two components. The
  spacing between the lines indicates an approximate estimate of the
  uncertainties on sidednesses and proper motion speeds. Sidedness
  ratio calculations are carried out with the form of the sidedness
  relation appropriate to a continuous jet. Note that only small
  angles to the line of sight ($\theta \sim 20\degr$) are consistent
  with all the observations.}
\label{constraints}
\end{figure*}

If we assume that the component speeds in the kpc-scale jet do
represent the bulk flow in the jet (and this seems inescapable in the
outer parts of the jet, where the moving features are extended) then
the combination of jet sidedness, assuming intrinsic jet symmetry, and
observed motion allows us to set some constraints on the speed and
angle to the line of sight of the jet. In general, the constraints
from jet sidedness and variability on VLBI scales have suggested that
Cen A is at a reasonably large angle to the line of sight ($\theta =
50\degr$ -- 80\degr, \citealt{tjrt98}): the small-scale jet-counterjet
ratio $R$ is between 4 and 8 \citep{jtmm96}, and the model-dependent
speed constraints ($\beta > 0.45$) then require large angles to avoid
obtaining $R > 8$. The situation is interestingly different in the
kpc-scale jet: $R > 50$ at knot A1B (where the limit comes from taking
the SJ3 component to be the brightest possible base knot counterpart),
and even in the A3 region $R \sim 10$ (the faint extended region S1 is
taken to be the counterpart of A3). These measurements, together with
the observed motions, give the constraints shown in Fig.\
\ref{constraints}. Taken at face value, they require much smaller
angles to the line of sight than has been estimated from the VLBI
observations: in fact, the angles to the line of sight proposed by
\citeauthor{tjrt98} are too large for any amount of beaming to be able
to account for the jet-counterjet ratio in the A1 region (Fig.\
\ref{constraints}), let alone the relatively mild beaming implied by
the observed sub-luminal proper motion. At least one of the
assumptions involved in the various estimates of $\theta$ must be
incorrect. Probably the jet and counterjet are not completely
intrinsically symmetrical, since we know that the larger-scale inner
lobes are asymmetrical both in their radio structure and their
environments \citep{kvfj03}: if this were true, some of the
constraints would be relaxed, particularly those based on the limits
on the sidedness near the compact A1 base knots, and it would be
possible to obtain consistent values of $\theta$ from the sub-pc and
100-pc-scale data. Angles to the line of sight $\sim 50\degr$ are
also more plausible in interpretations of the large-scale radio and X-ray
properties of the source \citep[e.g.,][]{kvfj03}. This interpretation
does leave unanswered the question of {\it why} the base region (A1) of the
northern jet is intrinsically much brighter than that of its presumed
southern counterpart.

The fact that the base knot A1A has apparently increased in flux
density by around 10\%\ can in principle give rise to some additional
constraints on speeds. A Gaussian fit to this feature suggests a FWHM
of about 0\farcs4, corresponding to 7.3 pc (23 light years). It is
therefore impossible for the knot to have varied coherently on a
timescale of a decade: it must contain some smaller-scale
substructure, which is consistent with the relatively small amplitude
of variation. Even so, the fact that it has changed so obviously
suggests that significantly relativistic speeds must be involved, at
least comparable to those estimated from proper motions further out.
If this is true, the fact that the knot is stationary does {\it not}
allow us to conclude that the bulk flow through the knot is slow.

\subsubsection{Polarization structure}

\begin{figure*}
\epsscale{0.85}
\plotone{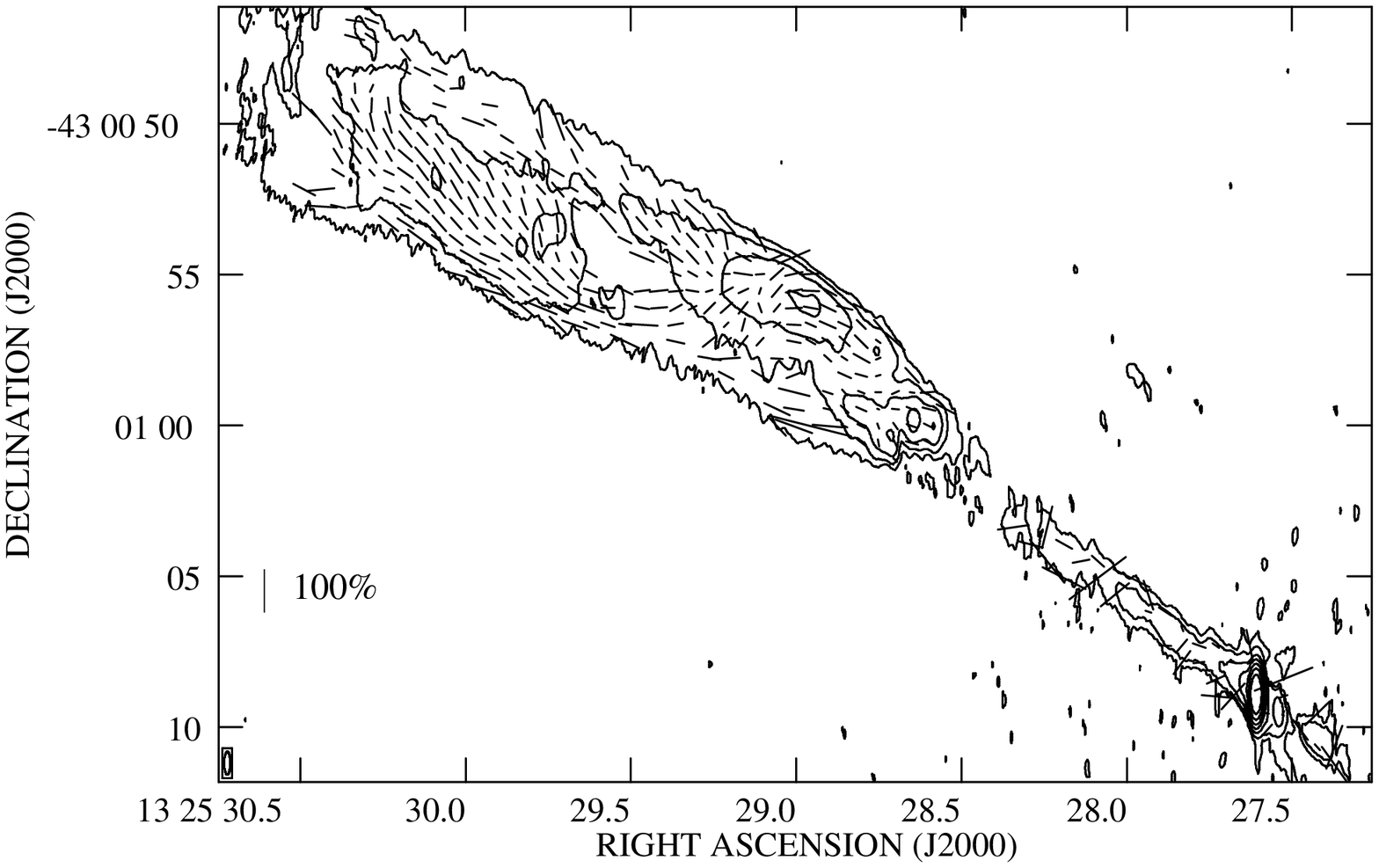}
\epsscale{0.85}
\plotone{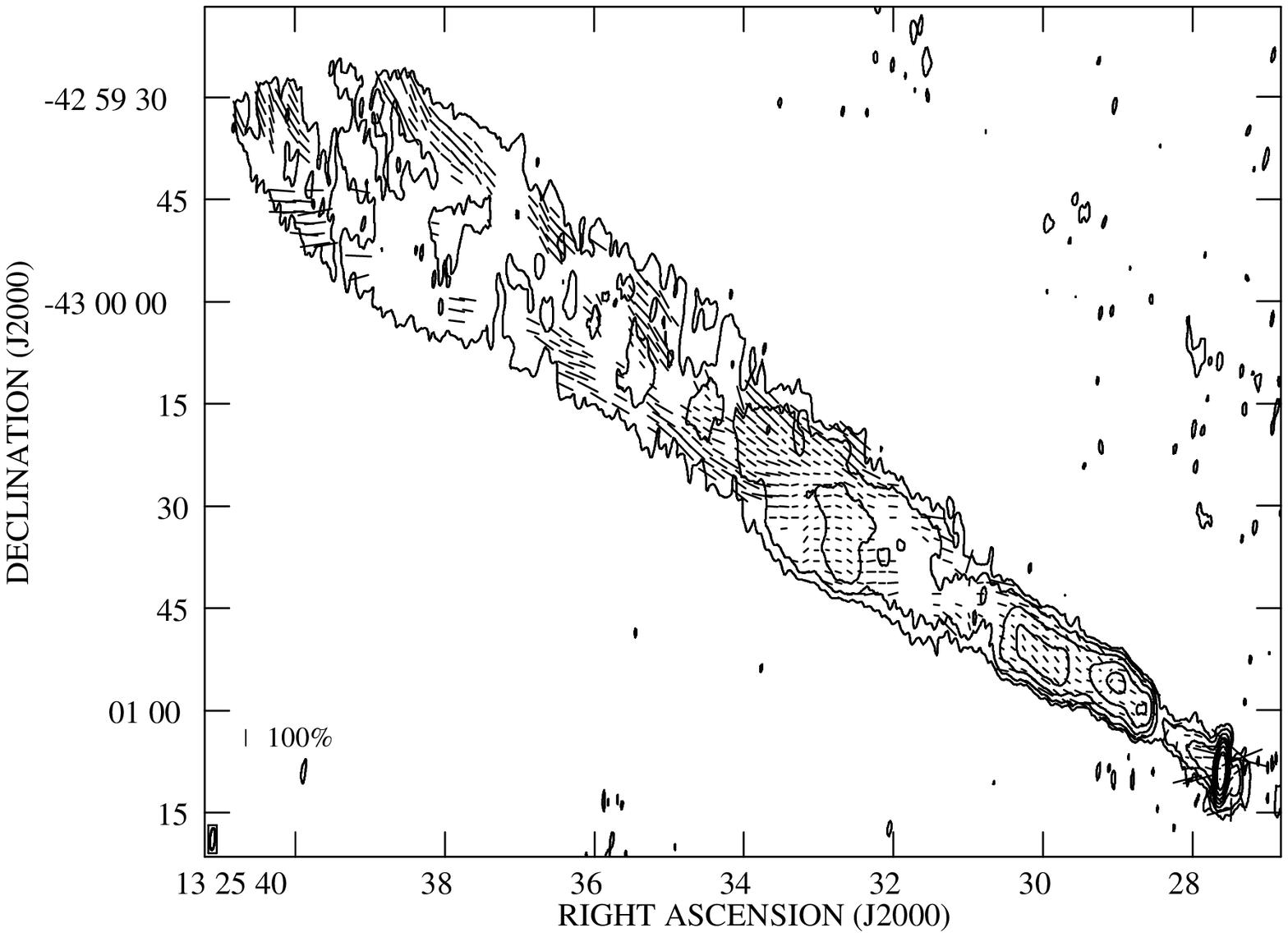}
\epsscale{1.0}
\caption{Polarization structure in the jet of Cen A. Top: the inner
  jet, image made from the 2002 A-array data with $0\farcs76 \times
  0\farcs20$ resolution (this image shows the region of the jet
  denoted by the outer box in Fig.\ \ref{radio}). Bottom: the
  larger-scale jet, image made from the 2002 B-array data with
  $3\farcs20 \times 0\farcs73$ resolution. Contours are at $200 \times
  (1,4,16,64,\dots)$ $\mu$Jy beam$^{-1}$ in both maps. The vector
  directions are perpendicular to the $E$-field direction, and so
  would show magnetic field direction if no Faraday rotation effects
  were present: the vector magnitudes show the relative degree of
  polarization. Note that (for simplicity of visualization) the
  vectors are uniform in RA and DEC, and so, given the elliptical
  beams, are oversampled in the N-S direction by a factor $\sim 3$.
  Vectors are only shown where the signal-to-noise in both total and
  polarized intensity maps exceeds $3\sigma$.}
\label{polar}
\end{figure*}

Our new observations give us good maps of the polarization structure
in the inner jet, and these are shown in Fig.\ \ref{polar}. They
confirm the basic picture seen in earlier observations \citep{bfs83,
cbf86} but have higher sensitivity and show more of the extended
structure of the jet. We plot vectors perpendicular to the
polarization $E$-vectors, which should be in the direction of the
magnetic field in the emitting material if Faraday rotation is
negligible, as we would expect it to be at this frequency from the
rotation measure images of \cite{cbn92}. The main point to note here
is that the magnetic field direction appears to be almost entirely
parallel to the jet over the whole region seen in these images (3 kpc
along the jet), with the exception of a small region to the E and S of
knot A2. There is no sign of the change to a transverse field
direction along the center of the jet that is seen in some other FRI
sources, and this appears to be the case on larger scales too
\citep{cbn92}, although we do not detect polarization along the
ridge-line of the large-scale jet. Accordingly, there is no evidence
that we are seeing a region of the jet where a parallel component of
the magnetic field at the edge might be thought to trace a slow-moving
shear layer, as in some proposed models of jet deceleration and
polarization structure \citep{l96}. More recent versions of these
models \citep[e.g.,][]{lb02a} do not predict a detectable transition
to a transverse field direction in all cases. It is also interesting
that the well-collimated inner part of the jet, before the flare point
at knot A, appears to show a parallel field structure. This has been
observed in some other sources \citep[e.g.,][]{ohc89,hapr96}.

\subsection{The radio/X-ray relation}
\label{RX}

\subsubsection{Identification of radio and X-ray features}

\begin{figure*}
\epsscale{0.9}
\plotone{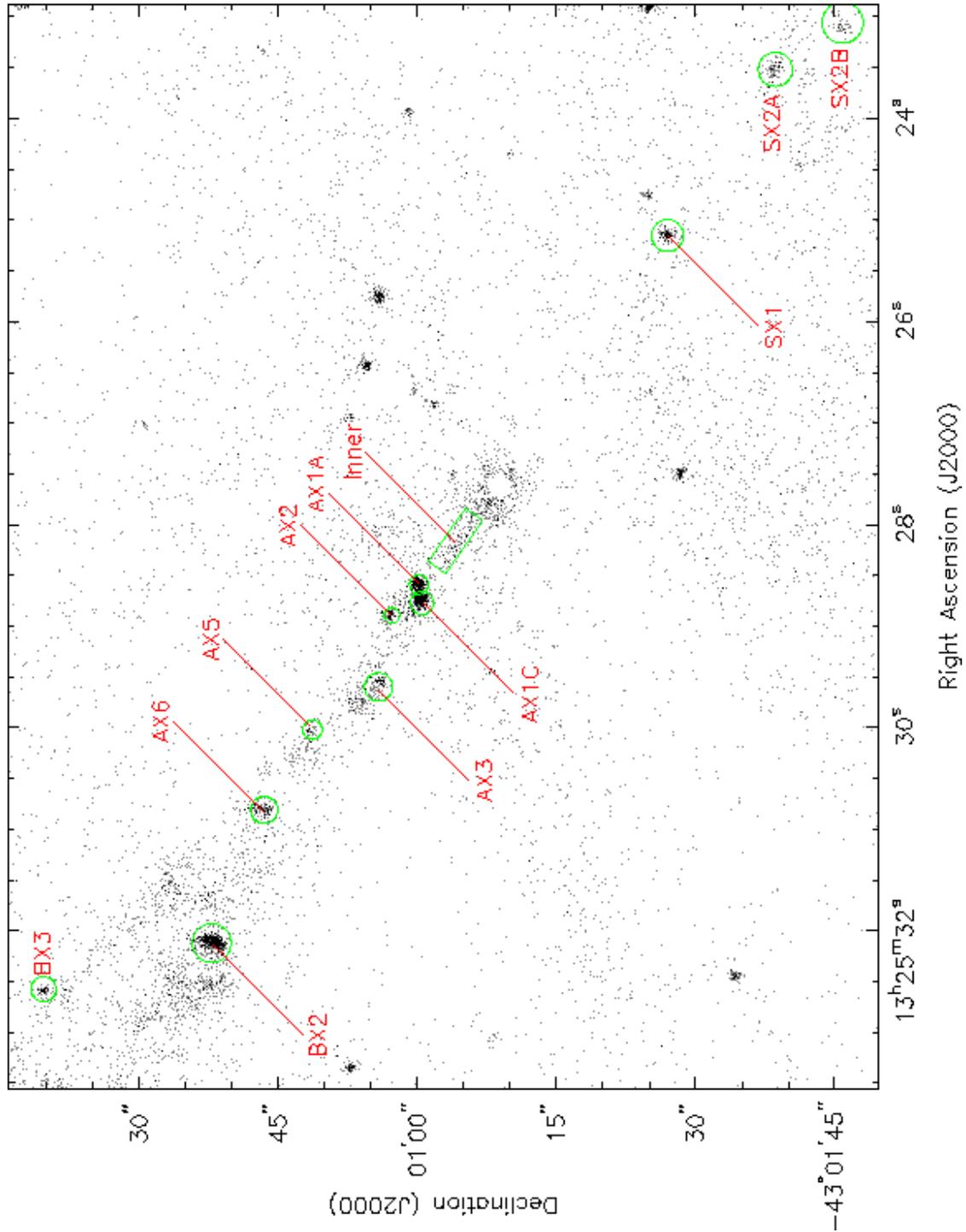}
\epsscale{1.0}
\caption{The X-ray emission from the Cen A and counterjet. Extraction
  regions for some of the X-ray features are labeled. The greyscale
  shows the raw counts in the energy band 0.4--2.5 keV, with
  $0\farcs123$ pixels. This image shows the
  same region of the jet as Fig.\ \ref{radio}.}
\label{extraction}
\end{figure*}

\begin{figure*}
\epsscale{0.9}
\plotone{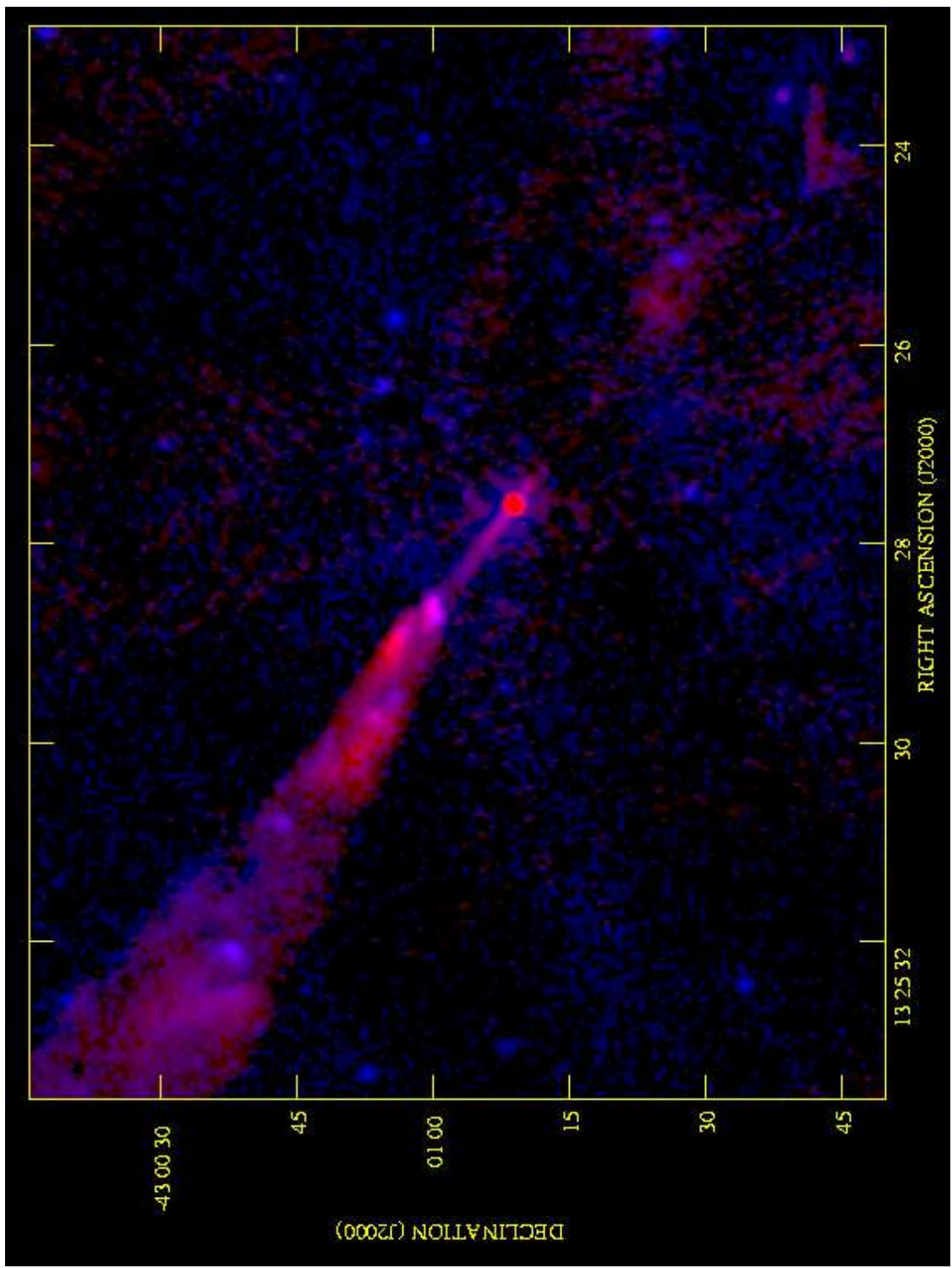}
\epsscale{1.0}
\caption{The X-ray and radio structure of the Cen A jet. The X-ray
  image, made from events with energies between 0.4 and 2.5 keV, is in
  blue and the radio image, made with {\it imagr} from the
  combined-epoch A- and B-configuration data, is in red. The X-ray
  data have been smoothed with a Gaussian and the restoring beam of
  the radio map has been chosen so that both radio and X-ray data have
  a resolution (FWHM) around $0\farcs 85$. The transfer function is
  non-linear to allow low-surface-brightness structure to be seen.
  This image shows the same region of the jet as Fig.\ \ref{radio}.}
\label{overlay}
\end{figure*}

The new X-ray and radio data allow a more detailed comparison to be
made between the two wavebands than was previously possible. The most
striking new results are the good detection (in X-rays and radio) of
the inner, well-collimated part of the jet, and the association of
several previously known X-ray knots with faint compact radio
features. Fig.\ \ref{extraction} shows the quality of the new X-ray
data, while Fig.\ \ref{overlay} shows a comparison of the radio and
X-ray structures. The unprocessed resolutions of the data are somewhat
different: the {\it Chandra} PSF close to the core can be fitted as a
circular Gaussian with FWHM $\sim 0\farcs65$, while the radio data
have an elongated beam, as discussed above. To simplify comparison in
Fig.\ \ref{overlay} we have smoothed the X-ray emission with a
$0\farcs5$ Gaussian kernel and used a circular restoring beam in the
radio mapping (after appropriate weighting of the $uv$ plane) so that
both images have a resolution of $\sim 0\farcs85$ (the FWHM of a
circular Gaussian) close to the core. The effective resolution of the
{\it Chandra} data is somewhat lower ($\sim 0\farcs95$) at the
furthest distances from the core shown on this Figure, but we do not
regard this as a significant problem.

It is clear from these images that some of the previously known X-ray
knots in the jet (AX2, AX3, AX5, AX6, BX2 in the notation of K02, as
used in Fig.\ \ref{extraction}) are associated with the newly
discovered weak radio knots: AX2 with A2A, AX3 with A3A, AX5 (or part
of it) with A5A, AX6 with A6A and BX2 with B1A. In addition, the two
X-ray knots SX2A and SX2B are coincident with the {\it counterjet}
radio features S2A and S2B. In the jet, extended X-ray emission is
associated with most, but not all, of the radio emission region: for
example, there is no strong X-ray emission associated with the bright
extended region A2 (as opposed to the weak compact knot A2A). In
several places there is brighter radio emission downstream of a faint
radio knot associated with an X-ray feature: this is true of the radio
features A2, A3 and B.

\begin{figure*}
\epsscale{1.0}
\plotone{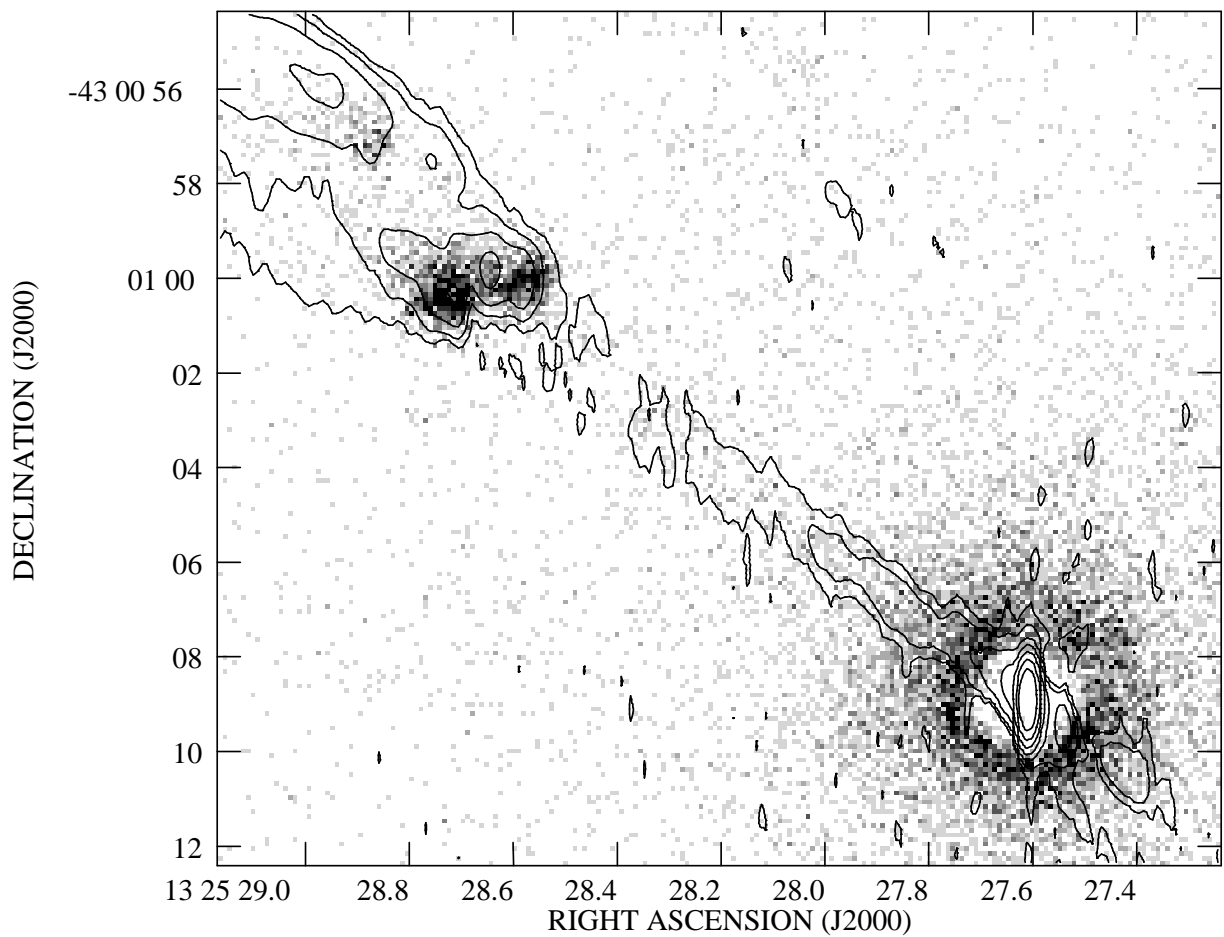}
\caption{The inner part of the Cen A X-ray and radio jet. The contour
  plot shows the A-array 2002 maximum-entropy map with resolution
  $0\farcs 76 \times 0\farcs 20$, with contours at $200 \times (1, 4,
  16, 64\dots)$ $\mu$Jy beam$^{-1}$, while the greyscale shows the
  0.5-7.0 keV X-ray data binned in 0\farcs0984 pixels (effective
  resolution $\sim 0\farcs65$). Black is 6 counts per pixel; the
  central parts of the X-ray nucleus are strongly affected by pileup
  and so no valid counts are seen. This image shows the region of the
  jet denoted by the inner box in Fig.\ \ref{radio}.}
\label{zoomover}
\end{figure*}

The results show that the radio to X-ray flux density ratio is
strongly variable as a function of position in the jet. Some radio
regions have comparatively strong radio emission and weak X-ray
emission (for example region A4) while others, of which the radio knot
B1A is probably the clearest example, have strong X-ray emission and
weak radio emission. There are some X-ray sources in or near the jet
which have no detectable radio counterparts. The X-ray feature BX3 is
an example of this, although it is possible that it is unrelated to
the jet (as suggested by its point-like X-ray appearance compared to
other jet sources: if it is unrelated to the jet, it is most probably
a LMXB associated with NGC 5128). Equally, there are comparatively
bright radio features with no X-ray detections. This is true of the
counterjet features SJ2 and SJ3 (Fig.\ \ref{overlay}). As Fig.\
\ref{zoomover} shows, there are also strong differences between the
X-ray properties of the knots in the bright A1 region (in this figure
we retain the full resolution of both datasets for clarity). The knot
A1B, the brightest of the three in the radio, has by far the faintest
X-ray emission. In the following subsection, we investigate the radio
and X-ray spectra of these regions quantitatively.

\subsubsection{X-ray spectroscopy and flux densities for compact features}

\begin{figure*}
\plotone{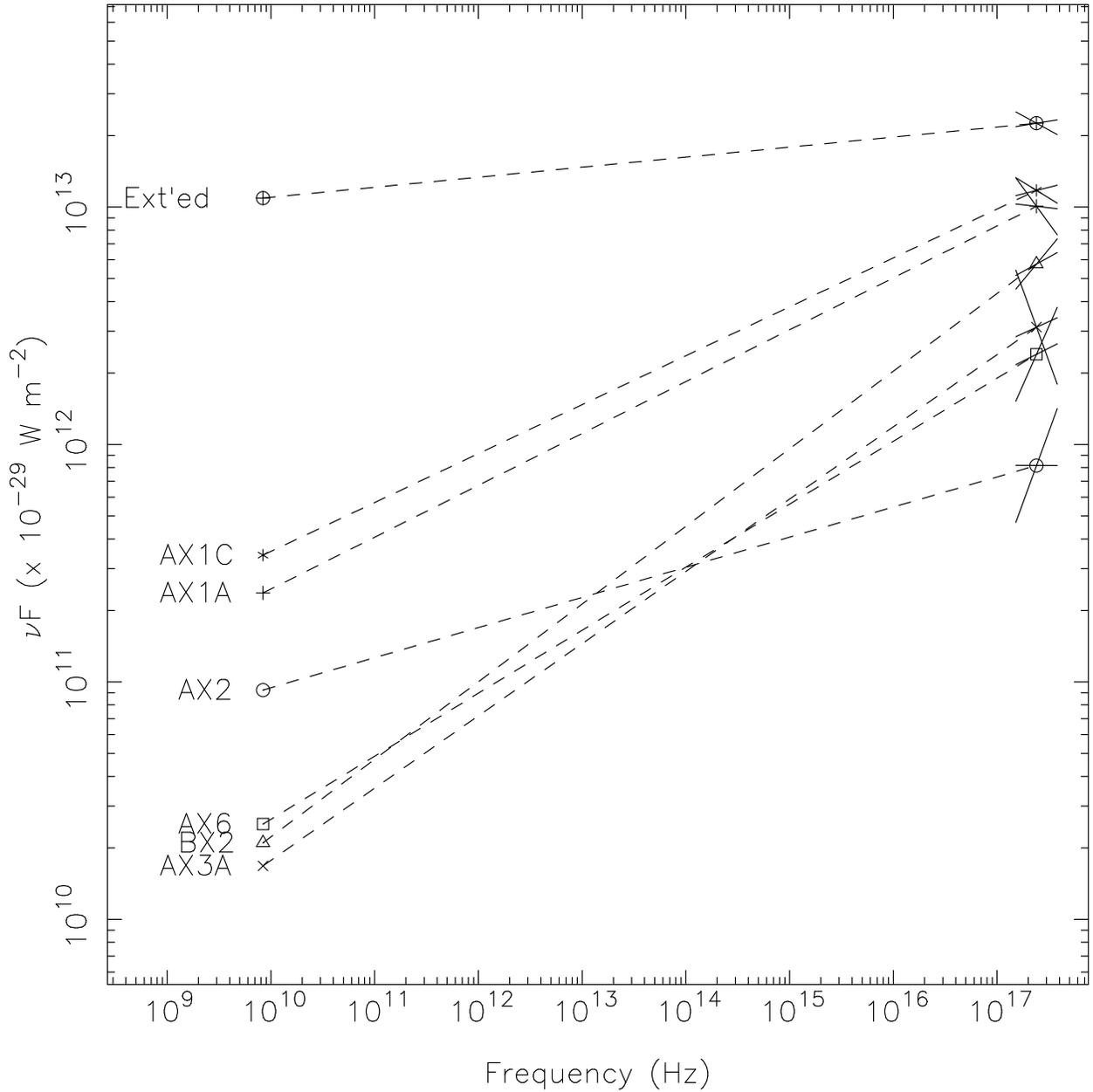}
\caption{The radio to X-ray spectra of some of the features of the Cen
  A jet, showing the range of X-ray to radio spectral indices (dashed
  lines) and X-ray spectral indices (`bow ties' around X-ray points).
  For clarity, the errors on the X-ray flux levels are not shown.}
\label{nufnu}
\end{figure*}

We extracted spectra and flux densities for all the compact X-ray
features associated with radio knots, together with the corresponding
radio flux densities. Fig.\ \ref{extraction} shows some of the
extraction regions, and the results are tabulated in Table
\ref{knots}; total counts in the knots vary between $\sim 1000$ for
the brightest features and $\sim 100$ for the faintest. In each case
we fitted a power-law model absorbed with a free, zero-redshift
absorbing column (which in general includes both the Galactic column
density, $\sim 7 \times 10^{20}$ cm$^{-2}$, and the much larger column
from the dust lane of Cen A). We included the effects of the reduced
quantum efficiency of the detector at the epoch of observations by
using the {\it acisabs} model to correct the response
files\footnote{As described in the {\it ciao} threads:
http://cxc.harvard.edu/ciao/threads/apply\_acisabs/ .}. The X-ray flux
densities quoted are the unabsorbed values. We determined background
using nearby, off-jet background regions. All the fits were good. The
corresponding radio flux densities are those of the compact features,
measured where possible by fitting a Gaussian and background level to
the radio images. We also tabulate, in parentheses, the {\it total}
radio flux densities (including flux from any extended background
emission seen in the A-array map) in the regions corresponding to the
X-ray extraction regions, which provides an upper limit on the radio
emission from the X-ray features. In some cases this total flux
density is a very conservative limit, as it includes part of another
radio feature: this is true of knots A1A and A1C (contaminated by A1B)
and A2A (contaminated by extended emission from the A2 region). We
find that the radio to X-ray flux ratios of the different knots vary
by more than an order of magnitude, and the best-fitting X-ray
spectral indices for radio-associated features range from 0.3 to 1.4.
Fig.\ \ref{nufnu} gives a graphical representation of the differences
in the properties of the knots.

The column densities inferred from the spectral fits, and tabulated in
Table \ref{knots}, are consistent with a single value of the absorbing
column, $\sim 6 \times 10^{21}$ cm$^{-2}$, in the inner regions of the
jet (AX1--6), as reported by K02. In knot BX2 and in the counterjet
features, the absorbing column density is lower, and is generally
consistent with the Galactic value. This is qualitatively as we would
expect from sensitive imaging of the dust features in Cen A
\citep[e.g.,][]{scms96}; the entire inner part of the jet lies in the
dust lane.

\begin{deluxetable}{llrrrrrr}
\tablecaption{X-ray and radio features of the jet}
\tabletypesize{\small}
\tablehead{X-ray name&Radio name&Radio flux&X-ray flux
  &$S_{\rm X}/S_{\rm R}$&$\alpha_{\rm RX}$&$\alpha_{\rm
    X}$&$N_{\rm H}$ ($\times $\\&&density (mJy)&density (nJy)&($\times
  10^{-6}$)&&&$10^{22}$ cm$^{-2}$)\\}
\startdata
AX1A&A1A&28.2 (74)&$41^{+8}_{-6}$&1.48 (0.56)&0.78&$1.30_{-0.26}^{+0.28}$&$0.61_{-0.13}^{+0.14}$\\
AX1B&A1B&51.2 (69)&$24^{+18}_{-11}$&0.47 (0.35)&0.85&$1.9_{-0.8}^{+1.0}$&$1.04_{-0.47}^{+0.60}$\\
AX1C&A1C&40.7 (113)&$47^{+6}_{-5}$&1.20 (0.43)&0.80&$1.05_{-0.18}^{+0.19}$&$0.47_{-0.08}^{+0.09}$\\
AX2&A2A&11 (50)&$3.2_{-0.5}^{+2.9}$&0.31 (0.07)&0.88&$0.1_{-0.3}^{+0.8}$&$0.004_{-0.004}^{+0.6}$\\
AX3A&A3A&2 (25)&$13^{+7}_{-4}$&6.5 (0.52)&0.70&$1.4_{-0.3}^{+0.7}$&$0.56^{+0.36}_{-0.29}$\\
AX6&A6A&3.0 (5.6)&$10^{+3}_{-2}$&3.3 (1.8)&0.73&$0.36_{-0.41}^{+0.42}$&$0.43_{-0.23}^{+0.29}$\\
BX2&B1A&2.5 (7.8)&$23_{-2}^{+2}$&9.7 (3.1)&0.68&$0.56_{-0.14}^{+0.15}$&$0.06_{-0.04}^{+0.04}$\\
BX3&--&$<1$&$2.3^{+0.7}_{-0.2}$&$>1.9$&$<0.76$&$0.2_{-0.5}^{+0.7}$&$<0.20$\\[5pt]
SX1&--&$<0.5$&$9^{+3}_{-2}$&$>18$&$<0.64$&$0.44_{-0.40}^{+0.42}$&$0.22_{-0.20}^{+0.22}$\\
SX2A&S2A&2 (2)&$3.2_{-0.5}^{+1.4}$&1.7 (1.7)&0.78&$0.45_{-0.40}^{+0.65}$&$0.04_{-0.04}^{+0.22}$\\
SX2B&S2B&2 (2)&$2.1_{-0.3}^{+1.0}$&0.9 (0.9)&0.80&$0.5^{+1.2}_{-0.7}$&$<0.33$\\[5pt]
Inner&Inner&28&$38_{-24}^{+98}$&1.43&0.79&$2.2_{-1.7}^{+2.5}$&$1.1_{-0.8}^{+1.2}$\\
Extended&Extended&1300&$86_{-5}^{+6}$&0.07&0.96&$1.00_{-0.15}^{+0.16}$&$0.11_{-0.03}^{+0.03}$\\
\enddata
\label{knots}
\tablecomments{Errors quoted for $\alpha_{\rm X}$ and $N_{\rm H}$ are
  the $1\sigma$ error for two interesting parameters ($\Delta \chi^2 =
  2.3$), since these two quantities are strongly correlated in the
  fits. For consistency, the limits quoted on the column density in
  the two cases where the best-fitting value is formally zero are also
  $1\sigma$ limits. The errors on the unabsorbed 1-keV flux densities
  are $1\sigma$ for one interesting parameter ($\Delta \chi^2 =
  1.0$).}
\end{deluxetable}

\subsubsection{Sizes of X-ray features}
\label{knotsize}

\begin{figure*}
\plottwo{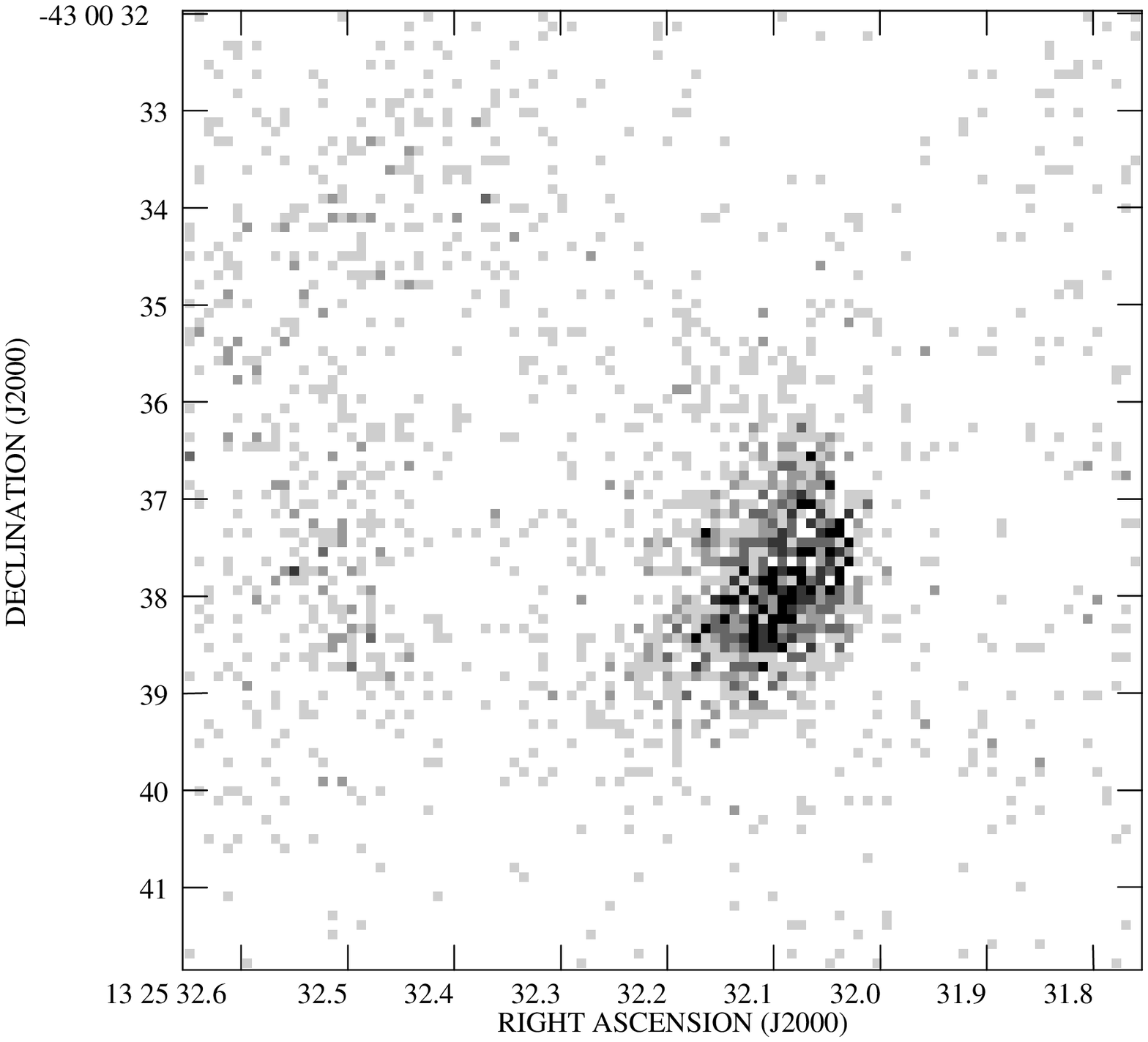}{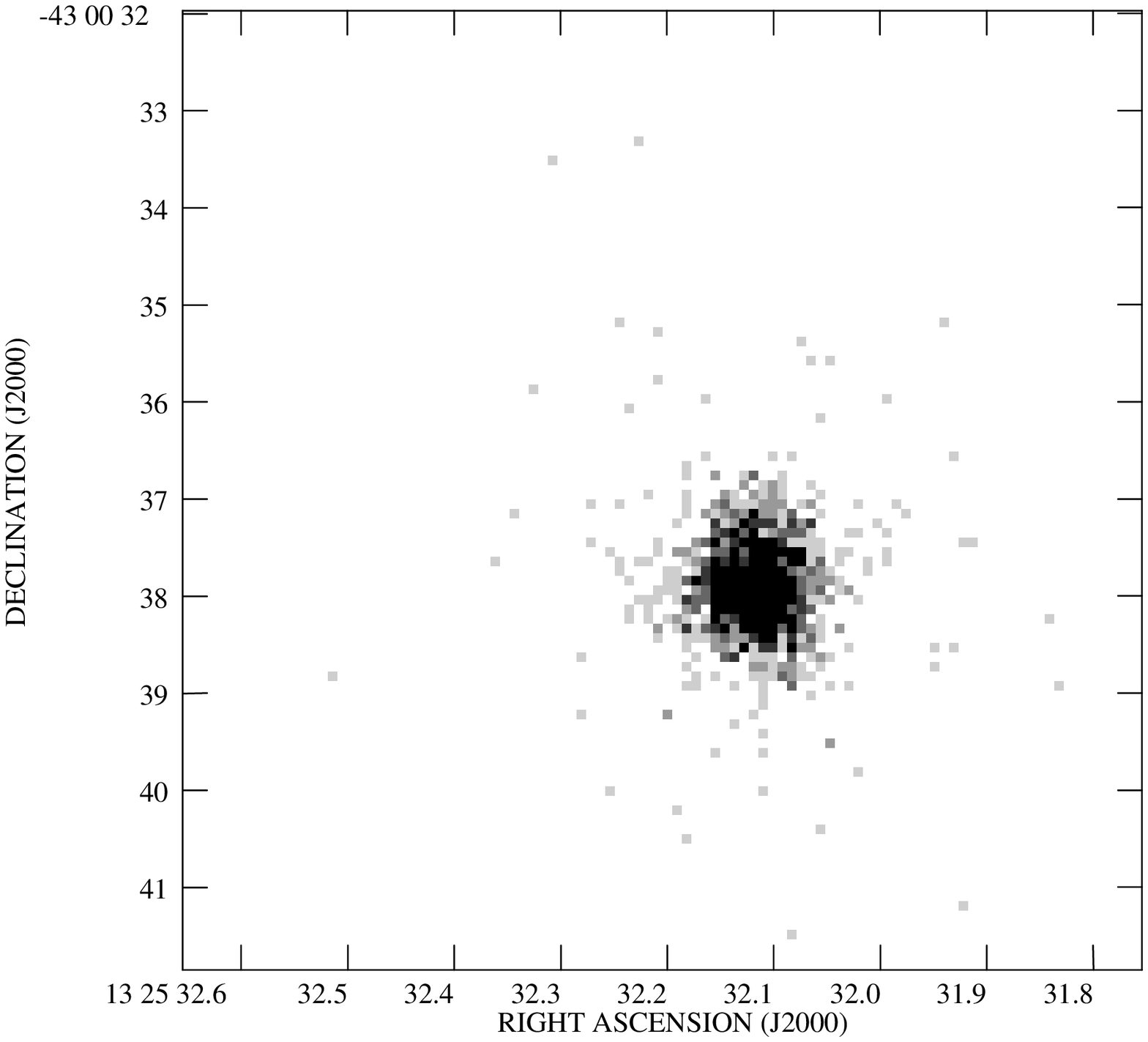}
\caption{The detailed structure of knot BX2 (left) and a simulated
  point-spread function normalized to have the same total counts
  within 2\arcsec\ (right). BX2 is clearly extended on scales of approximately 1\arcsec.
  One pixel is $0\farcs0984$ and black is 5 counts per pixel. See the
  text for details of the simulations used.}
\label{bx2}
\end{figure*}

We noted in K02 that several of the X-ray knots in the jet appeared to
be significantly extended. Some of these (the most obvious example
being AX1: Fig.\ \ref{zoomover}) are now seen to have sub-structure on
scales smaller than the resolution of the data that was then available
to us. Knot BX2, by contrast, appears to be genuinely resolved in our
new data. Fig.\ \ref{bx2} shows a comparison between the observations
and a simulation, using the web-based ray-tracing tool {\it
ChaRT}\footnote{See http://cxc.harvard.edu/chart/ ; we followed the
threads described at http://cxc.harvard.edu/chart/threads/prep/ and
http://cxc.harvard.edu/chart/threads/marx/ in order to generate
simulated point sources matched to our data.}, of the {\it Chandra}
point-spread function at this position for the observed energy
distribution of BX2. Visual inspection and radial profiling both show
that BX2 is resolved {\it transversely} to the jet direction, on
scales of 1\arcsec. This extension is also present in the radio data,
though the elongated radio beam makes it less obvious. We have
searched for X-ray extension in the other isolated, compact
radio-related knots such as AX6 and found little significant evidence
that it is present, which implies sizes in the X-ray $\la 0\farcs5$
($\la 10$ pc); these are again consistent with the radio observations.
The complex structure of the AX1 region means that a full
deconvolution, which we have not carried out, would be required to
derive good constraints on structure in the X-ray sub-knots, but we
believe that both bright X-ray knots are probably marginally extended
on scales $\sim 0\farcs5$.

\subsubsection{The extended X-ray and radio emission}

We extracted two spectra for extended regions of the jet. These were
the faint inner jet between the nucleus and knot A1, and the extended
emission in the jet between knot A2 and the knot B region, excluding
all compact X-ray features. In both cases we used an off-source
background region of the same size at the same radial distance from
the nucleus. The radio emission was measured from the corresponding
regions of the multi-epoch, A+B-configuration images without
background subtraction. The results are tabulated in Table
\ref{knots}. Note that the X-ray to radio flux ratio for the extended
jet is much lower than for any individual jet knot.

\begin{figure*}
\plotone{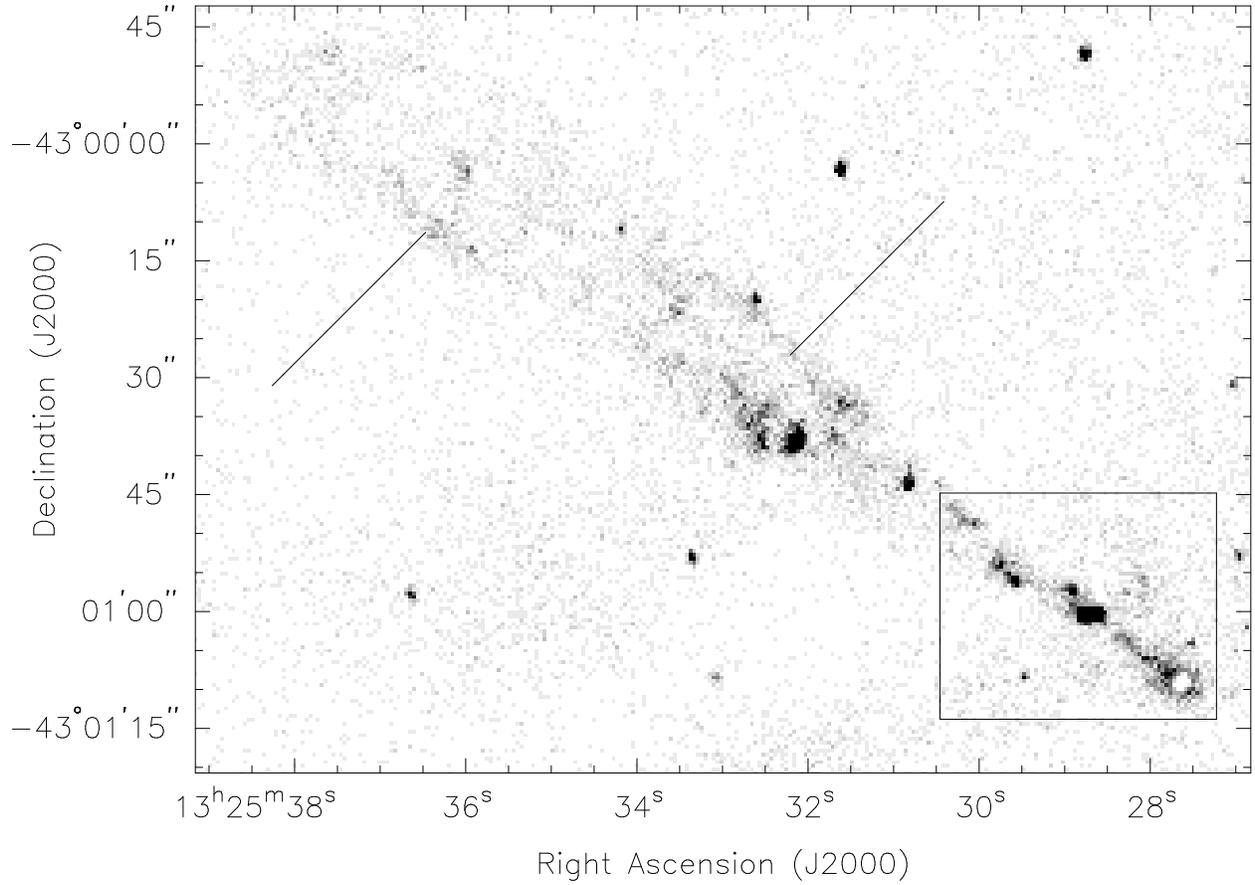}
\caption{The X-ray structure of the large-scale Cen A jet. The
  greyscale shows the raw counts in the 0.4--2.5 keV band; the pixels
  are $0\farcs492$ on a side and black is 12 counts per pixel.
 The lines indicate two
  regions where the X-ray jet appears significantly edge-brightened
  (see the text). The box, shown for orientation purposes only, shows
  the region denoted by the outer box in Fig.\ \ref{radio}. }
\label{largejet}
\end{figure*}

The X-ray emission on scales larger than that of the inner jet is
shown in Fig.\ \ref{largejet}. Generally the extended X-rays are
reasonably well matched to the radio emission on these scales; in
particular, diffuse X-ray emission clearly extends to the edge of the
radio jet. There are some faint compact X-ray features, identified by
K02, which have no apparent compact radio counterparts, though we
cannot rule out the possibility that their radio counterparts are
simply too faint to be convincingly detected. To the north and
downstream of the bright X-ray knot BX2 the extended X-rays appear to
be associated with the edge of the jet, and to be absent in the jet
center. A similar region of edge-brightening is seen further down the
jet. These regions are marked with lines on Fig.\ \ref{largejet};
neither has a counterpart in the large-scale radio emission.
Their detection raises the possibility that the edges of the jet,
which are the locations where jet mass entrainment takes place in the
standard FRI jet model \citep[e.g.,][]{lb02a} are privileged
sites for the particle acceleration required to generate the X-ray
emission.

\section{Discussion}

\subsection{Radio supernovae?}

Although the X-ray knots with detected radio counterparts cannot be
X-ray binaries in Cen A, since such objects have at most extremely
weak radio counterparts, it is not out of the question that they could
be jet-associated radio supernovae (SN) or supernova remnants (SNR).
\cite{c02} has suggested that in some cases type Ia supernovae may be
triggered by jets\footnote{\cite{c02} also suggested that SN1986G in
NGC 5128 was associated with the jet in Cen A, but in fact the
position of the SN means that this is not the case (Capetti, private
communication, 2002). The location of SN1986G, a type Ia SN, is not at
present a detectable source of radio or X-rays.}. X-ray emission of
luminosity comparable to that of the X-ray knots in Cen A (K02) is
most likely to be produced by type II supernovae in dense environments
rather than by SNIa, but there is evidence \citep{gf02} that the jet
is inducing star formation in the host galaxy, so that this is not
impossible. The radio luminosities of the weaker knots, with fluxes of
a few mJy, are comparable to those of known SNR in M82 \citep{mpwa94},
which lies at a very similar distance, and at least some of the M82
sources have tentative X-ray associations of similar luminosity to the
Cen A knots \cite{mtka01}, although a full analysis of the M82 {\it
Chandra} data has not yet been published. The main fact that convinces
us that the Cen A knots are not related to supernovae is their good
power-law spectra. The emission from a young SNR in a dense
environment is expected to be thermal and to have a comparatively low
temperature, $\la 1$ keV \citep[e.g.,][]{ft96} while the only thermal
models that can be fitted to flat-spectrum sources like BX2 require
high temperatures ($>6$ keV at the 90\% confidence level) and very low
metal abundance. We therefore assume in what follows that the radio
and X-ray knots reflect structures in the fluid flow in the jet.

\subsection{Knot properties}

We can rule out the possibility that the X-ray bright, radio-faint
knots are simply compressions in the synchrotron-emitting fluid that
makes up the extended jet. As an example, we consider the case of knot
BX2, which has a well-constrained, flat $\alpha_{\rm RX}$ and
$\alpha_{\rm X}$. The extended jet has a steep $\alpha_{\rm RX}$ and
$\alpha_{\rm X}$, and so, if we model its spectrum as a simple broken
power law in frequency, with a low-frequency (radio) spectral index
similar to the $\alpha_{\rm RX}$ of BX2, the break to a steeper spectral index
must occur at comparatively low frequencies, $\nu_{\rm b} \sim
10^{12}$ Hz. For adiabatic compression of this material to produce the
spectrum seen in knot BX2, where the break frequency would be above
the soft X-ray band ($\nu_{\rm b} \ga 5 \times 10^{17}$ Hz), we
require 1-dimensional compression factors $\cal R$ of about 30, since
$\nu_{\rm b} \propto {\cal R}^{-4}$ for a tangled field geometry
\citep{l91}. But such high compression factors would increase the
radio volume emissivity of the knot over that of the parent material
by a factor $\sim 10^{11}$ ($j_\nu \propto {\cal R}^{-(5+4\alpha)}$),
whereas the ratio of the volume emissivities of the knot and extended
jet (assuming spherical symmetry for the one and a truncated-cone
geometry with uniform filling factor for the other) is $\sim 2$. BX2
is an extreme case, but similar arguments apply to the other compact
jet features.

Instead, it must be the case that the knots are privileged sites for
the {\it in situ} particle acceleration that is required throughout
the jet. For the the base knots (AX1A, AX1C) this is not particularly
surprising; these are presumably related to the transition between the
faint, well-collimated, efficient inner jet and the much brighter
extended jet (Fig.\ \ref{zoomover}). They can be modeled as standing
shocks at the base of the jet, and we cannot even rule out the
possibility that their radio-to-X-ray spectra are described by a
standard continuous injection model \citep[e.g.,][]{l91}, as used to
describe the hotspots in FRII radio sources, in which the spectral
index steepens from $\sim 0.5$ to $\sim 1.0$ at some frequency.

What does this imply for the weaker knots --- AX2, AX3A, AX6, BX2 and
the counterjet features? One clue is provided by the fact that none of
the radio counterparts of these features appears to be moving with the
jet flow, although for some of them (AX2, AX3A, AX6) there is nearby
and downstream apparently moving structure. This strongly suggests
that these features are also standing shocks in the jets, which is
consistent with the observation that at least one of them is extended
perpendicular to the jet direction (\S\ref{knotsize}). It is very hard
to imagine a model purely related to the fluid flow in the jet that
would give rise to this kind of localized stationary shock after the
jet flare point at knot A1. Instead,
the shocks must be related to some feature of the jet's environment
that is fixed or slowly-moving in the galaxy frame, a point we return
to below (\S\ref{stars}).

By contrast, the features that are clearly moving in the radio images
(A1B, A2, A3/A4) have comparatively little X-ray emission; A2 and A4
in particular appear to have X-ray to radio ratios less than the
values typical for the extended jet as a whole. For the speeds
inferred from the proper motion and sidedness constraints (Fig.\
\ref{constraints}) the Doppler factor is $>1$ in these regions, so
beaming, for a fixed spectral shape, would tend to {\it increase} the
observed X-ray to radio flux density ratio (since the spectral index
of the X-ray-emitting material is steeper than the radio spectral
index, so that the $K$-correction is greater). So it is likely that the
high-energy particle acceleration in these regions is less efficient
than in the jet as a whole. This would again be consistent with a picture in
which the distributed particle acceleration process is more efficient
in the slower-moving edges of the jet than in the faster-moving
regions which we would expect to be closer to the jet center.

The detection of some large-scale counterjet X-ray features with
associated radio knots confirms that there is continuing energy
transport, and thus presumably collimated outflow, on scales of $50''$
or 850 pc, projected, from the core on the counterjet side of
Cen A. It is still not clear whether the 6 or so X-ray features
without radio counterparts that lie in the inferred counterjet region
(Fig.\ \ref{overlay}) are all related to the counterjet. The
properties of the brightest of these, SX1 (Table \ref{knots}), are
extreme compared to those of the jet knots, but we cannot rule out a
flat-spectrum power law ($\alpha \sim 0.5$) extending from the radio
to the X-ray, which would give rise to a radio flux density around 50
$\mu$Jy; our radio images are still not quite sensitive enough to
detect this.

\subsection{Standing shocks}
\label{stars}

If the radio-weak, X-ray bright, stationary or slow-moving knots in
the post-A1 region are indeed standing shocks in the jet, then we
would expect them to be related to some feature of the jet environment
approximately fixed in the frame of the host galaxy. Here we
concentrate on models in which the jet is interacting with discrete,
compact objects \citep{bk79}. We can obtain a constraint on the mass
of these objects from the fact that they are not observed to move. The
minimum energy density $\epsilon$ in the jet is of order a few $\times
10^{-11}$ J m$^{-3}$. If we assume that the bulk flow speed $v_j \sim
0.5c$ (i.e. only mildly relativistic, with Lorentz factor $\sim 1$),
then a limit on the mass $M$ of an object not moving visibly with the
flow is given by

\[
M \ga \pi R^2 \epsilon \left({v_j\over c}\right) {t\over v_l}
\]
where $v_l$ is the minimum speed of proper motions that we could
observe (say $0.1c$), $R$ is the radius of the object, and $t$ is the
length of time that the object has been experiencing the thrust from
the jet; we assume that the kinetic energy density of the jet
dominates its mass-energy density, as is the case for a plasma
consisting only of relativistic electrons. Considering the minimum
energy in the inner lobes ($E \sim 2 \times 10^{48}$ J) and requiring
that this should have been supplied by the jet, $E \approx \epsilon
A_X v_j t$ where $A_X$ is the cross-sectional area of the jet, we
obtain $t$ of the order of a few $\times 10^6$ years, which we can
treat as a limit on the timescale that the jet has been currently
active (clearly the total energy in the middle and outer lobes of Cen
A is much larger, but these may well be the results of earlier epochs
of AGN activity.) Taking $R$ to be comparable to the sizes of the
smaller radio knots, $\sim 10$ pc, and using a jet radius of 60 pc, we
obtain values of $M$ for the obstacle of the order of a few solar
masses (or more). $M$ can be less than this if $R$ or $t$ are less
than our estimates. Given the velocity dispersion in NGC 5128
\citep{i98}, the time for an individual star to cross the jet would be
of order a few $\times 10^5$ years, which would relax the limit on $M$
somewhat. On the other hand, there is no particular reason to believe
that the energy density in the jet is the minimum energy, and some
evidence that it is greater for FRI jets \citep[e.g.,][]{lb02b} which
would increase the required mass.

We expect that there will be $\sim 10^9$ stars in the inner kpc of the
galaxy (the region including the compact knots) and so the jet would
be expected to include a few $\times 10^6$ stars at any given time:
even if stars are ablated rapidly by the jet, new stars would enter
from the edges on short timescales. Clearly the particle acceleration
regions cannot be associated with individual normal stars, although it
is possible that `shocklets' around each star contribute to the
required diffuse high-energy particle acceleration which occurs
throughout the jet.

The objects responsible for the discrete radio and X-ray features
(AX2, AX3A, AX6, BX2 and the counterjet features) must be considerably
rarer than normal stars. Possibilities include high-mass-loss stars
(e.g.\ Wolf-Rayet stars) and entrained gas clouds. The size of the
interaction region $r_w$ for a high-mass-loss star is given by ram-pressure
balance between the stellar wind with speed $v_w$ and the jet:
\[
r_w = \left({\dot M v_w c^2\over 4\pi\epsilon v_j}\right)^{1\over2}
\]
and so, for a Wolf-Rayet with a mass-loss rate $\dot M$ of
$10^{-4}M_\odot$ yr$^{-1}$ and $v_w \sim 2000$ km s$^{-1}$ we can
readily obtain scale sizes of the order of 10 pc if the jet is near
its minimum energy density. These are rather extreme properties even
for a Wolf-Rayet, however \citep[e.g.,][]{nl00}, and would require
what is a rather large (though not impossibly large) number of
Wolf-Rayets per normal star, given the stellar densities estimated
above; the Wolf-Rayet fraction is known to depend strongly on the star
formation rate.

On the other hand, various types of gaseous material are present in
the inner part of the galaxy. Hot gas is known to be present from its
thermal X-ray emission, but its central density is only $3.7 \times
10^{-2}$ cm$^{-3}$ \citep{kvfj03} which means that a 10-pc-diameter
cloud would be a factor of a few below our derived lower mass limit.
Molecular material is known to be present in the inner part of Cen A,
particularly the dust lane \citep[e.g.,][]{we00}: for a density of
$\sim 300$ cm$^{-3}$, a 10-pc-diameter molecular cloud would have $8
\times 10^3\ M_\odot$, which is certainly not excessive given the
estimated total molecular hydrogen mass of $4 \times 10^8\ M_\odot$
\citep{i98}, and more than satisfies the constraints on mass derived
above. $10^4$-K line-emitting material, which is known to be
associated with the inner jet \citep*{bkb83} as well as being present
on larger scales \citep{mrfd91} would have similar or somewhat lower
densities to molecular material, and would be equally suitable as
obstacles if the clump size was great enough; alternatively, the line
emission seen by \citealt{bkb83} could be the result of stripping and
shock-excitation of colder material. In any case, we consider
interaction with clouds of cold or warm gas to be more likely than
interaction with high mass-loss stars as the explanation for the
stationary radio and X-ray knots in Cen A.

Finally, it should be borne in mind that Cen A's jet is probably an
order of magnitude lower in kinetic luminosity than the jets in
well-known 3C FRI sources with X-ray jets like 3C\,31 and 3C\,66B.
This means that the stellar-wind interaction model, at least, is
probably not viable as an explanation for any X-ray/radio features in
those jets. Interaction with external gas clouds is still a possibility,
but Cen A's host is probably richer in molecular material than a
typical FRI host elliptical.

\subsection{The nature of `offsets'}
\label{offsets}

The data for Cen A emphasize the importance of high spatial resolution in
discussing apparent offsets between the radio and X-ray peaks in FRI
jets. The strong variation in X-ray to radio flux ratio as a function
of position that we observe in Cen A could give rise to apparent
offsets between the peaks of unresolved or poorly resolved
knots in more distant sources. For example, if Cen A were placed at
the approximate distance of a well-studied nearby FRI like 3C\,66B
($z=0.0215$), the resolution of {\it Chandra} and the VLA would
correspond to tens of arcsec on Fig.\ \ref{overlay}. The knot B region
would then be essentially unresolved, and would present a classical
example of an offset between the radio and X-ray peaks, entirely
because of the very different $\alpha_{\rm RX}$ values of BX2 and the
remaining downstream regions of knot B. In order to make an adequate
model of the physics underlying the `offset' behavior, it is necessary
to have radio and X-ray data with spatial resolution sufficient to
sample the jet structure on the physical scales of interest.
Unfortunately, Cen A is at present the only source for which that is
true.

\begin{figure*}
\plottwo{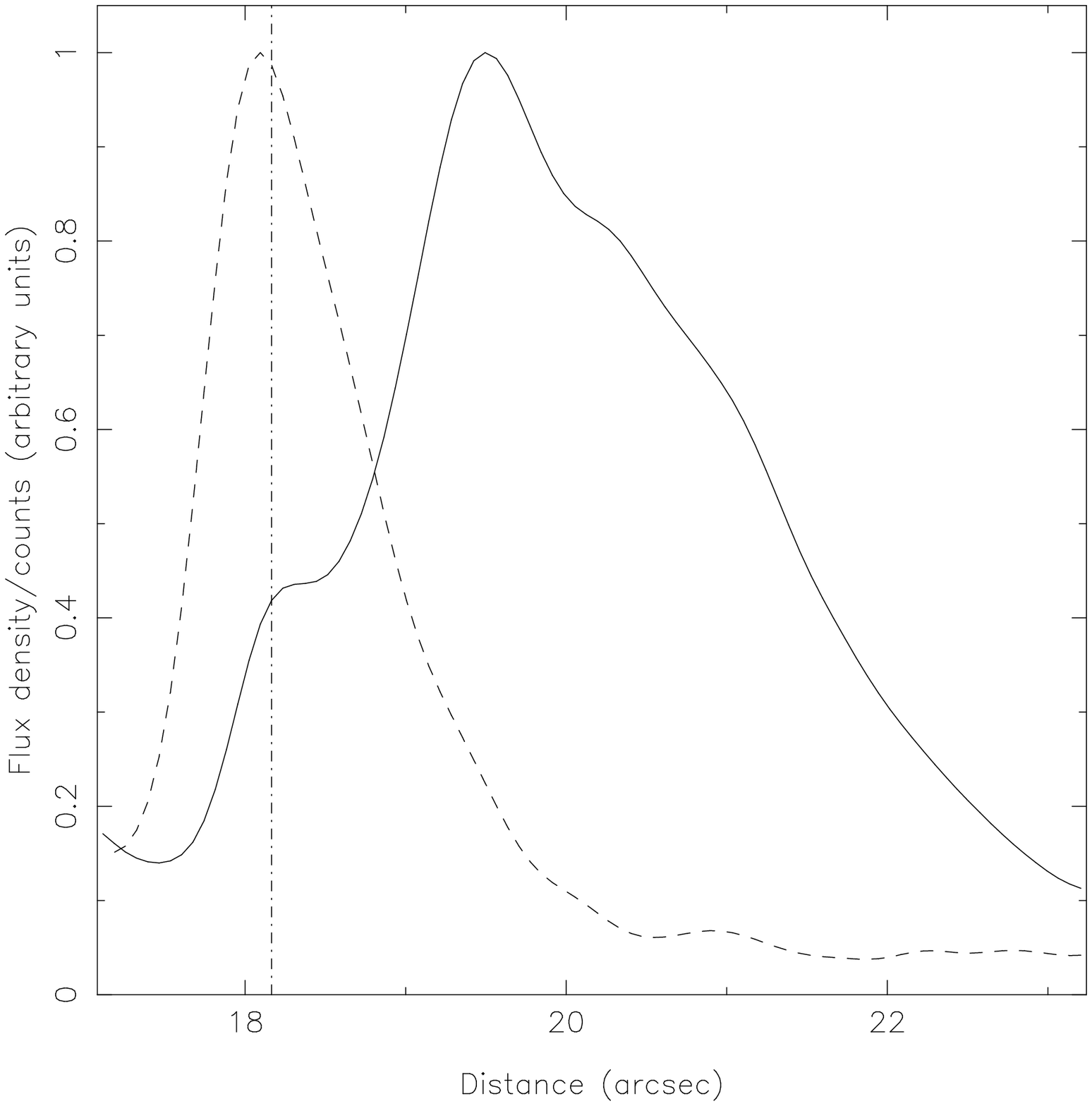}{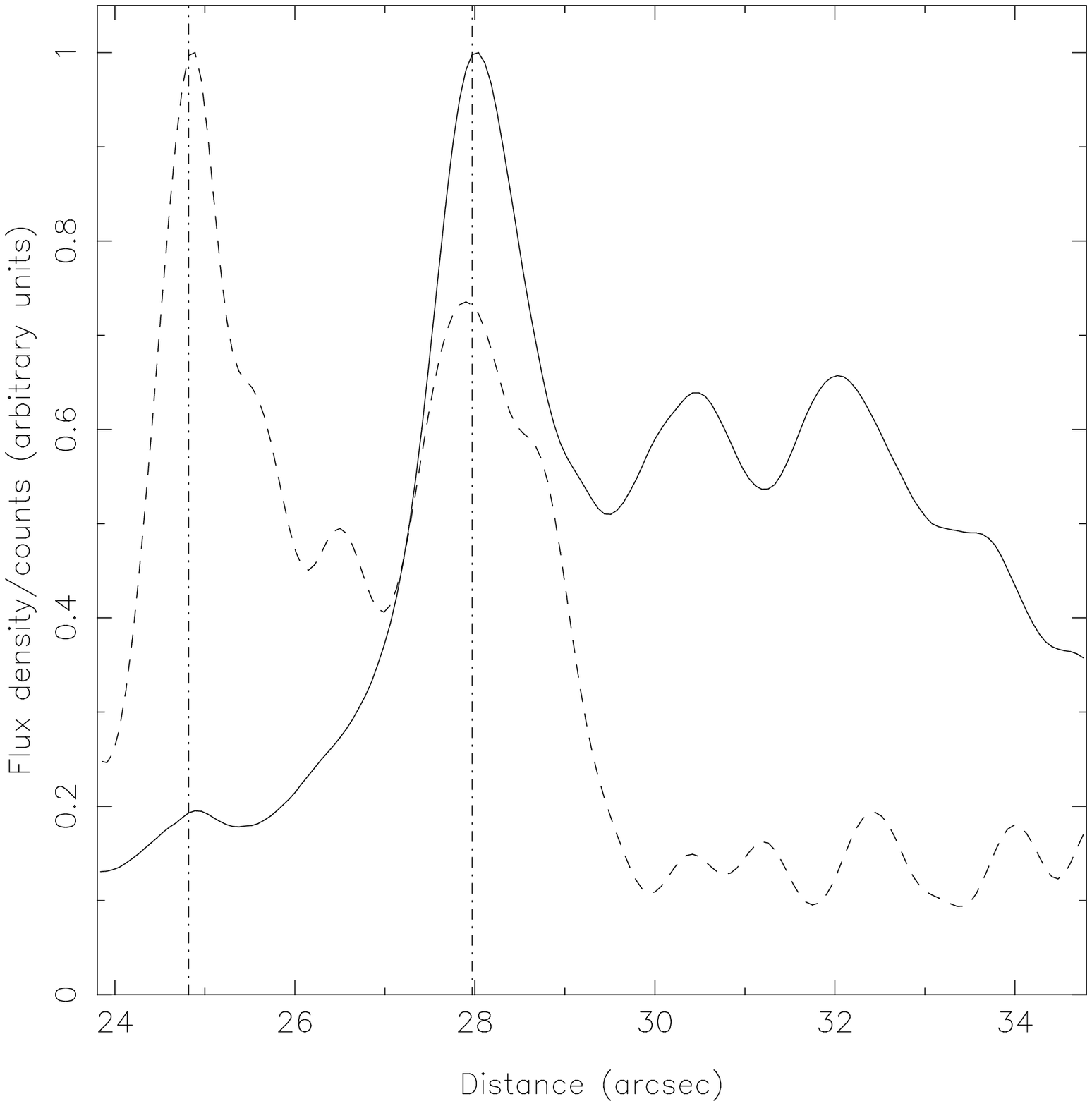}
\caption{Radio and X-ray profiles along the jet in the regions of the
  radio features A2 (left) and A3/4 (right). The solid line shows the
  radio profile taken from the full-resolution 2002 A-configuration
  image, while the dashed line shows the profile of the X-ray emission
  (convolved with a Gaussian of FWHM $0\farcs5$). Vertical dot-dashed
  lines show the locations of the compact radio knots A2A (X-ray knot
  AX2), A3A (AX3A) and A3B (AX3B). The distance plotted on the
  $x$-axis is the distance from the nucleus, measured along the center
  of the jet. The normalizations of the X-ray and radio data have been
  re-scaled for ease of viewing in both plots. Note the differences in
  position between the peaks of the X-ray and radio emission in both
  profiles.}
\label{slices}
\end{figure*}

It is still of interest to ask why the observed `offsets' are always
in the sense that the X-ray peak lies closer to the nucleus. We
suggested \citep{hbw01} that the offsets in knot B of
3C\,66B could be modeled in terms of a particle-accelerating shock
together with downstream advection of radio-emitting particles, while
the X-ray-emitting electrons would be rapidly quenched by synchrotron
losses and/or expansion (a model subsequently discussed in more detail
by \citeauthor{bl03} [\citeyear{bl03}]). Of the features in Cen A, region A2 comes
closest to this simple picture (Fig.\ \ref{slices}). Almost all the
X-ray emission comes from a region coincident with a very faint
compact radio knot (A2A); downstream there is bright radio emission
with little or no corresponding X-ray. The X-ray profile is certainly
what we would expect from a model with particle acceleration and
downstream advection. But this model does not simply explain the clear
separation ($1\farcs3$, or 20 pc) between the faint knot A2A,
coincident with the X-ray peak, and the peak of the radio emission. If
the downstream advection were uniform, then, for a static region of
particle acceleration, we would expect the radio profile to be
brightest at the same place as the X-ray, and then to fade more slowly
as a function of downstream distance. This
model would explain the observed offsets in other, more distant
sources, but it is {\it not} consistent with what we actually see in
Cen A. If we require all the radio-emitting particles in the A2 region
to have been accelerated at A2A, then we need, at least,
non-uniform downstream advection to cause them to `pile up' at the A2
peak.

There are two other areas in the inner jet where bright radio emission
is seen downstream of a faint radio/bright X-ray knot, in the regions
of knots A3/4 and B. These depart even more clearly from the expected
behavior in the simple downstream advection model. Fig.\ \ref{slices}
shows that the weak radio knot A3A, coincident with the brightest
X-ray emission, leads the radio peak (A3B) by about 3\arcsec\ (50 pc).
However, in this case, there is X-ray emission from the radio peak
too. In knot B, the X-ray peak, coincident with the radio knot B1A,
is separated by around 8\arcsec\ (140 pc) from the peak of the
downstream diffuse emission, which again shows some weak X-ray
emission.

We cannot rule out the possibility that the apparent association
between weak upstream radio knots with bright X-ray emission and
bright downstream diffuse radio features with faint X-ray emission is
coincidental. There are only three clear cases, and other knots (such
as A6 and the counterjet features) do not have bright downstream
emission. However, if it is not coincidental, then it is certainly
giving us information about the properties of the fluid flow. For
example, if the compact knots are standing shocks in the flow, caused
by interaction with static obstacles (\S\ref{stars}) then it is
conceivable that the bright downstream radio emission is due to
compression and/or turbulent particle acceleration as the
post-obstacle flow re-joins the main jet. In this case the downstream
distance $d$ and the size of the obstacle $R$ would give us an
estimate of the internal Mach number of the jet, ${\cal M} \approx
d/R$. The knot-to-peak distances for A2A and A3A are respectively
about 20 and 50 pc, which would lead to estimates of ${\cal M}$ of
approximately 2 and 5 respectively for our adopted obstacle size of
$\sim 10$ pc. (As the peak downstream distance for knot B is not well
defined, we do not include it here.) These Mach numbers are at least
in the expected region for the base of an FRI jet, being mildly
supersonic, and (given our proper motion results) would imply
relativistic internal sound speeds in these regions of the jet.

Finally, we note that offsets between the peak positions of radio and
X-ray knots have been seen in the other extragalactic jet where
comparatively high spatial resolution is available, M87's
\citep{wy02}, where 1\arcsec = 78 pc. The peak-to-peak distances in
the M87 knots that show offsets are also tens of parsecs, consistent
with what we find above for the knots in Cen A, although the M87 jet
is clearly rather physically different from Cen A's (being somewhat
narrower and much smoother) in the regions where the offset is
observed. Based on our results above, we would predict that X-ray
knots D and F in M87 are actually coincident with faint, as yet unseen
features in the radio jet. If this is the case, the good optical
information available for M87's jet should place interesting
constraints on the knot spectra.

\subsection{The inner jet}

The detection of the well-collimated inner jet in both radio and
X-rays is of interest because, in the standard model of FRI radio
sources, this represents the efficient, supersonic flow that
transports energy up to the point where the jet disrupts and becomes
trans-sonic and turbulent (which would be at knot A in Cen A). These
jets are efficient in the sense that their radio luminosity per unit
length is much less than is observed after the flare point. However,
the detection of X-ray emission from these regions (in addition to Cen
A, the inner jets of M87 [\citealt{wy02}] and 3C\,66B
[\citealt{hbw01}] have been detected in the X-ray) is of interest
because, if it is synchrotron in origin, it shows that this region of
the jet is still capable of {\it in situ} particle
acceleration.\footnote{The situation is altered if the emitting
regions of the inner jet are highly relativistic, with a bulk Lorentz
factor $\Gamma \sim 10$, as inferred for some FRII jets. In this case,
the particle lifetimes are increased due to time dilatation. However,
since, for plausible angles to the line of sight, the Doppler factor
for such a jet would be less than 1, the observed X-rays would be
generated by higher-energy electrons than for a sub-relativistic jet,
while the inferred jet-frame magnetic field strength would be
increased; at the same time, the energy density of microwave
background and, more importantly, galactic photons in the jet frame is
increased by a factor $\sim \Gamma^2$, shortening the jet-frame loss
timescale of the X-ray emitting electrons to the inverse-Compton
process by the same factor. If the observed X-rays are in fact
synchrotron, it is hard to evade the necessity for {\it in situ}
acceleration. Producing the X-rays via the inverse-Compton process
would require very large departures from equipartition, very large
bulk Lorentz factors ($\Gamma \sim 50$) coupled with small angles to
the line of sight, or a combination of the two.} In M87 and 3C\,66B
the inner jet has a considerably higher X-ray to radio ratio than the
regions further from the nucleus. In Cen A the situation is
complicated by the poorly constrained absorbing column in this region,
which means that the unabsorbed flux density of the jet is uncertain
(Table \ref{knots}) but it is still clearly the case that the X-ray to
radio ratio is higher than that in the extended jet as a whole, though
probably not higher than those of some of the compact knots. (Here we
assume, since the spatial resolution is insufficient to allow us to do
anything else, that the X-ray and radio emission from the inner jet
come from the same regions.) Unlike the M87 and 3C\,66B inner jets,
Cen A's shows little evidence in radio or X-ray for knotty
sub-structure which might be associated with oblique internal shocks
giving rise to particle acceleration, while other possible particle
acceleration mechanisms (such as second-order Fermi acceleration or
turbulent magnetic field line reconnection) would be expected to be
more efficient in the trans-sonic large-scale jet. Doppler boosting
may be important here, particularly given the steep best-fitting X-ray
spectral index. The observed X-ray to radio ratio goes as ${\cal
D}^{(\alpha_{\rm X} - \alpha_{\rm R})}$ and so, to bring the ratio in
the inner jet in line with that in the extended jet (assuming little
Doppler boosting in the latter), we need an inner jet Doppler factor
$\sim 6$ (with a large uncertainty due to the uncertainty in the inner
jet flux density and spectral index). Doppler factors of this order
would require $\theta \la 10\degr$, which is more or less possible
given the constraints of Fig.\ \ref{constraints}, though, as we
suggest above, such small angles to the line of sight are unlikely for
other reasons. Nevertheless, we cannot rule out the possibility that
rapid bulk motion in Cen A's inner jet is responsible for its high
X-ray to radio flux ratio. If this were the case, the radio emission
from the jet as we observe it would be boosted by a large factor, so
that in reality it would be still more efficient in transporting
energy.

\section{Summary and conclusions}

\begin{enumerate}
\item Our high dynamic range radio images of Cen A reveal subluminal
apparent motions ($v \sim 0.5c$) in the hundred-parsec scale jet. Some
extended regions of the jet appear to be moving coherently downstream,
which suggests that the apparent speeds may be close to the bulk jet
speed. If this is the case, and if the jet and counterjet are
symmetrical, Cen A must make a comparatively small angle to the line
of sight ($\theta \sim 15\degr$), which contrasts with the larger
angles inferred from the parsec-scale properties of the source.
\item We have discovered faint radio counterparts to a number of the
previously unidentified X-ray knots in the inner parts of the jet,
demonstrating that the X-ray features are jet-related.
\item  We also detect
radio counterparts to some X-ray features on the counterjet side,
suggesting that there is collimated flow on kpc scales in the
counterjet region.
\item If the X-rays from the compact knots are due to
synchrotron emission, then the radio to X-ray and X-ray spectra allow
us to rule out a model in which the knots are simply compressions in
the fluid flow: instead, they must be privileged sites for high-energy
particle acceleration.
\item Almost all the strongly X-ray emitting knots appear to have
radio counterparts that are static within the limits of our
observations, suggesting that they trace stationary shocks in the jet
flow. Plausibly they are the result of an interaction between the jet
fluid and an internal obstacle such as a high-mass-loss star or
molecular cloud. By contrast, the radio jet features that are
apparently moving show weak or absent X-ray emission, although there
is still diffuse X-ray emission throughout the jet that is not
identified with discrete radio features.
\item Several of the radio-faint, X-ray-bright knots are associated
with downstream bright radio emission, and we suggest that it is this
behavior, seen at lower resolution, which gives rise to the observed
offsets between the radio and X-ray peaks in some more distant FRI
jets. The simplest possible model with particle acceleration and
downstream advection does not explain the details of these observations.
\end{enumerate}
\acknowledgments

We are grateful to Mark Birkinshaw for discussion of the jet fluid
dynamics, and to an anonymous referee for a number of comments which
helped us improve the presentation of the paper.

MJH thanks the Royal Society for a research fellowship.

The National Radio Astronomy Observatory is a facility of the National
Science Foundation operated under cooperative agreement by Associated
Universities, Inc.

\clearpage

\end{document}